\documentclass[10pt]{article}

\usepackage{amsmath}
\usepackage{amssymb}

\usepackage{graphicx}

\usepackage{cite}

\usepackage{color} 

\usepackage{url}

\usepackage[labelfont=bf,labelsep=period,justification=raggedright]{caption}
\usepackage{subfig}

\makeatletter
\renewcommand{\@biblabel}[1]{\quad#1.}
\makeatother

\date{}

\begin{document}

\begin{flushleft}
{\Large
\textbf{A universal law in human mobility}
}
\\
Xiao Liang, 
Jichang Zhao, 
Ke Xu$^{\ast}$
\\
State Key Lab of Software Development Environment, Beihang University, Beijing 100191, P.R. China
\\
$\ast$ E-mail: kexu@nlsde.buaa.edu.cn
\end{flushleft}

\section*{Abstract}
The intrinsic factor that drives the human movement remains unclear for decades. While our observations from intra-urban and inter-urban trips both demonstrate a universal law in human mobility. Be specific, the probability from one location to another is inversely proportional to the number of population living in locations which are closer than the destination. A simple rank-based model is then presented, which is parameterless but predicts human flows with a convincing fidelity. Besides, comparison with other models shows that our model is more stable and fundamental at different spatial scales by implying the strong correlation between human mobility and social relationship.


\section*{Introduction}
In recent years, there are increasing available human trajectories collected by diverse ``sensors'', which provide us opportunities to gain deeper insight into human mobility patterns. Specifically, there are some empirical studies on human mobility with the help of distinct ``sensors'' for track data acquisition such as banknotes \cite{Brockmann2006}, cellular towers \cite{Gonzalez2008,Song2010a,Kang2012,Toole2012}, GPS \cite{Jiang2009,Bazzani2010,Liang2012} and online location-based social services \cite{Cheng2011,Noulas2011a,Cho2011}. Though these studies have greatly improved our understanding of human mobility patterns, there is still lack of a unified framework to explain and model human mobility. First, studied mobility datasets collected by different acquisition techniques often capture different population groups' movements at different spatial scales or resolutions in different geographical environments. Therefore, some inconsistent results are observed. For example, heavy-tailed \cite{Brockmann2006, Gonzalez2008} and exponential \cite{Veloso2011,Bazzani2010,Kang2012,Roth2011,Liang2012,Peng2012} distributions of human trip lengths are observed at the intra-urban and inter-urban levels respectively. Some research thinks the exponential law is emerged in trips captured by a single means of transportation while the scaling law in movements by various kinds of transportation modes \cite{Szell2011,Yan2013}. However, another study attributes the different laws to different geographical population distributions at different spatial scales \cite{Liang2013}. Second, many agent-based mobility models \cite{Brockmann2006,Song2010a,Han2011,Hu2011a,Jia2012} are proposed to model individual movements, which try to reproduce some features observed at the aggregated level. But the diversity of individual mobility patterns is discovered recently \cite{Yan2013} and a general understanding of human movements is still missed. Third, for collective movements, the gravity model  \cite{Barthelemy2010} is widely used to predict human flows \cite{Balcan22122009, Jung2012, Goh2012}. But the parameters in the model must be fitted from actual flux in advance. In addition, the model cannot explain the asymmetry of traffic flows between any two locations. Then the radiation model \cite{Simini2012a} is proposed to deepen our understanding of human mobility, which is nonparametric and suitable for predicting commuting and migration flows. However, it is restricted by movements datasets \cite{Masucci2013} and is incompetent to characterize human daily movements at the intra-urban level \cite{Liang2013,Yan}. Because of this, there are new studies aiming to model human flows in cities \cite{Liang2013,Yan}. To be brief, all the above reasons make putting forward a general model for both intra-urban and inter-urban mobility necessary, which is exactly the aim of the present work.

In this paper, we explore human travels at different spatial scales and 
try to unify the understanding of general human mobility patterns.
First of all, a universal law for human mobility is demonstrated at intra-urban and inter-urban levels. Then based on the universal law we propose a rank-based mobility model that is parameter-free, enabling us to predict human flows only depending on geographical population distribution. Finally, we compare some recent mobility models and suggest that the rank-based model is more simple, stable and fundamental. Most of important, it could bridge human mobility and social relationship intrinsically.

\section*{Results}
\subsection*{A universal law}
In this paper, we define the rank of location $l$ relative to location $o$ as the total population number of locations which are closer to the location $o$ than the location $l$. Specifically, the rank $rank_o(l)$ is formulated as follows
\begin{gather}
rank_o(l) = \sum_{i\in S}{P_i}, \\
S = \{i|d(o, l_i) < d(o, l), i=1,\cdots,N\}, \nonumber
\end{gather}
where $l_i$ represents the $i$th location with the population $P_i$, $N$ is the total number of locations and the function $d(\cdot, \cdot)$ stands for the geographic distance between two locations. At the same time, supporting the spatial distances between any two persons chosen from two specified locations separately are equal, then $rank_o(l)$ is consistent with the general definition of the rank of any person $u$ in the location $l$ with respect to any person $v$ in the location $o$ \cite{Liben-Nowell2005}. 

Therefore, a trip from one location to another can correspond to the rank of the pairwise locations. Assuming $T(r)$ is the total number of trips with the same value of rank $r$, the ratio $P(r)$ can be defined as
\begin{equation}
P(r) = \frac{T(r)}{\sum_{rank_{l_i}(l_j)=r}{P_{l_i} \cdot P_{l_j}}}.
\label{eqn:rank}
\end{equation}
The denominator is the total number of pairs of people with rank $r$ and it suggests the potential likelihood of movements with rank $r$. So $P(r)$ is proportional to the average probability of travels with rank $r$.
\begin{figure}[htbp]
\centering
\includegraphics[scale=.4]{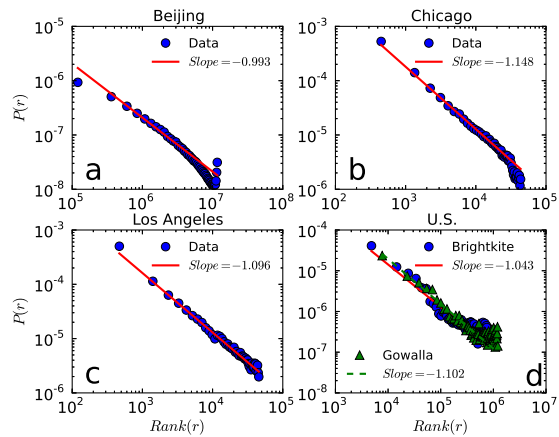}
\caption{{\bf The relationships between the ratio $P(r)$ and the rank $r$ in the four datasets.} (a)Beijing. (b)Chicago. (c)Los Angeles. (d)U.S..}
\label{fig:move_rank}
\end{figure}

As shown in Fig. \ref{fig:move_rank}, the relationships between $P(r)$ and $r$ in the four datasets all can be well approximated by power-law distributions with power exponents approaching -1.0. That is to say, the ratio $P(r)$ is inversely proportional to the rank $r$. It is worthy noting that our mobility datasets come from different sources, including taxis, travel surveys and location-based services. Furthermore, the datasets of Beijing, Chicago and Los Angeles characterize human movements at the city scale, while the dataset of U.S. at the country level. Though the diversity of these datasets, the same rule $P(r) \propto 1/r$ still holds for human mobility, which means the probability of a trip is inversely proportional to the number of population closer to the origin than the destination. The universal law illustrates that the rank is more appropriate than the distance to characterize human movements because it integrates the non-uniformity of population density naturally.

\subsection*{The rank-based model}
Inspired by the discovery of rank-based mobility, we can model human flows between different locations. Given the original location $i$ and the destination location $j$ with population $P_i$ and $P_j$ separately, the rank $rank_i(j)$ is the total number of people who live closer to $i$ than people in the location $j$. In terms of Equation (\ref{eqn:rank}), the flux $T_{ij}$ from the location $i$ to $j$ should satisfy 
$$\frac{T_{ij}}{P_i\cdot P_j} = \frac{T_i\cdot P(j|i)}{P_i\cdot P_j} \propto \frac{1}{rank_i(j)}$$
where $T_i$ is the total number of trips originating from the location $i$ and $P(j|i)$ is the probability of a trip to the location $j$ given the starting location $i$. Then $P(j|i)$ can be normalized as 
$$P(j|i) = \frac{P_j/rank_i(j)}{\sum_{k\neq i}{P_k/rank_i(k)}}$$
Ultimately, the traffic flows $T_{ij}$ is concluded as
\begin{equation}T_{ij}=T_i\frac{P_j/rank_i(j)}{\sum_{k\neq i}{P_k/rank_i(k)}}.\label{eqn:prediction}\end{equation}
When considered range of human mobility is fine-grained corresponding to relative large number of locations, the denominator of Equation (\ref{eqn:prediction}) can be approximated as
$$\sum_{k\neq i}{P_k/rank_i(k)}\approx \sum_{n=1}^{M-1}{1/n} \approx \ln M,$$
where $M$ is the total number of population. Hence the prediction of human flows is simplified as
$$T_{ij}\approx T_i\frac{P_j}{rank_{i}(j)\cdot \ln M}.$$
This avoids the summation for all locations in the denominator as in Equation (\ref{eqn:prediction}) and makes the calculation of traffic flux very simple while without losing accuracy too much.

Recently, the parameter-free radiation model is proposed and it can give competitive prediction for commuting or migration flows at the country level \cite{Simini2012a, Masucci2013}. But some studies reveal that the model may be incompetent to characterize more frequent intra-urban mobility \cite{Masucci2013, Liang2013, Yan}. It must be noted that our datasets record human movements with variety purposes, which are not restricted to commuting or migration. And the regularity of rank-based mobility has been demonstrated in these datasets. So in this paper, we focus on more general human movements and intend to use the law of rank-based mobility to predict human flows between zones at the city level or regions at the country level.

\begin{figure*}[htbp]
\centering
\includegraphics[scale=.18]{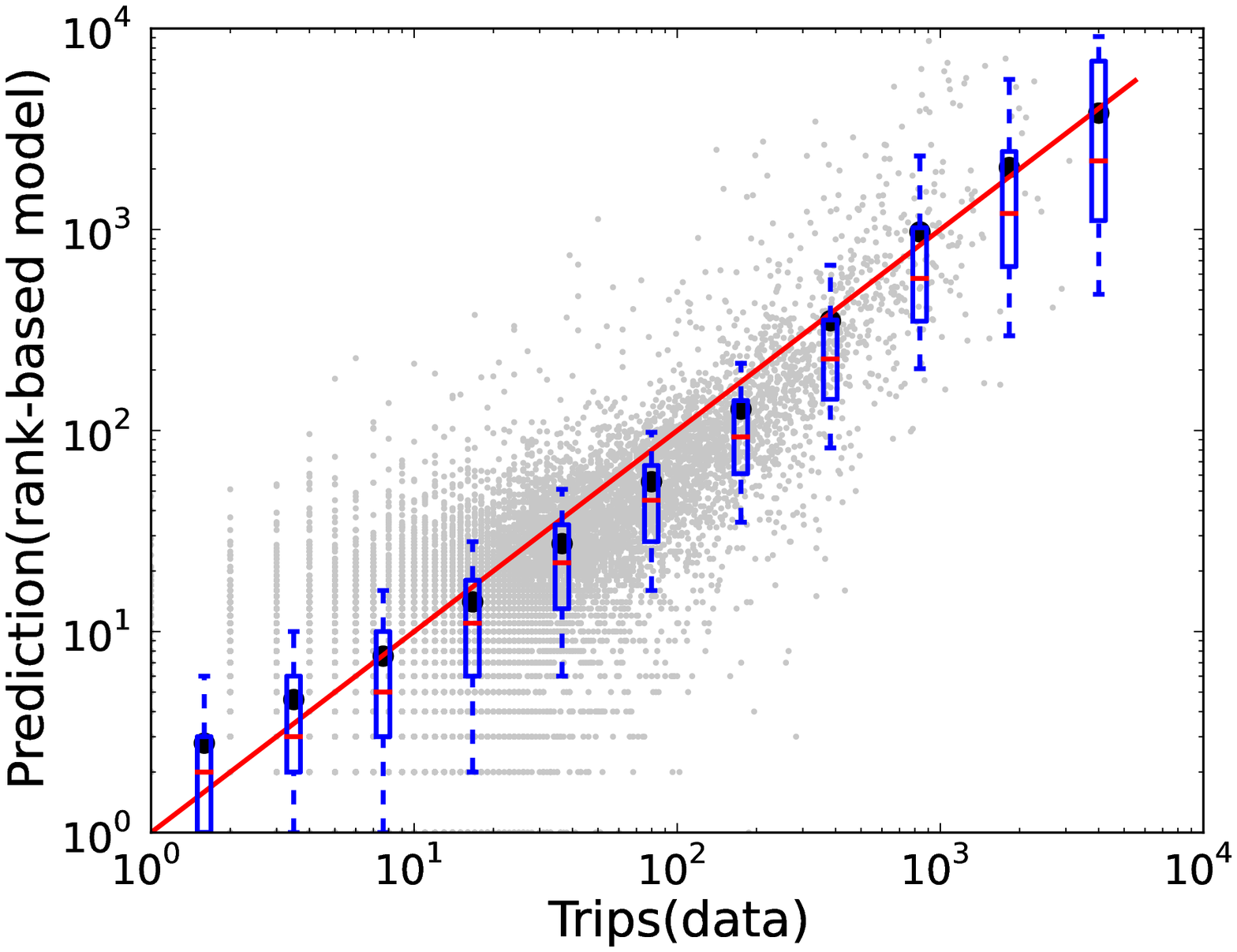}
\includegraphics[scale=.18]{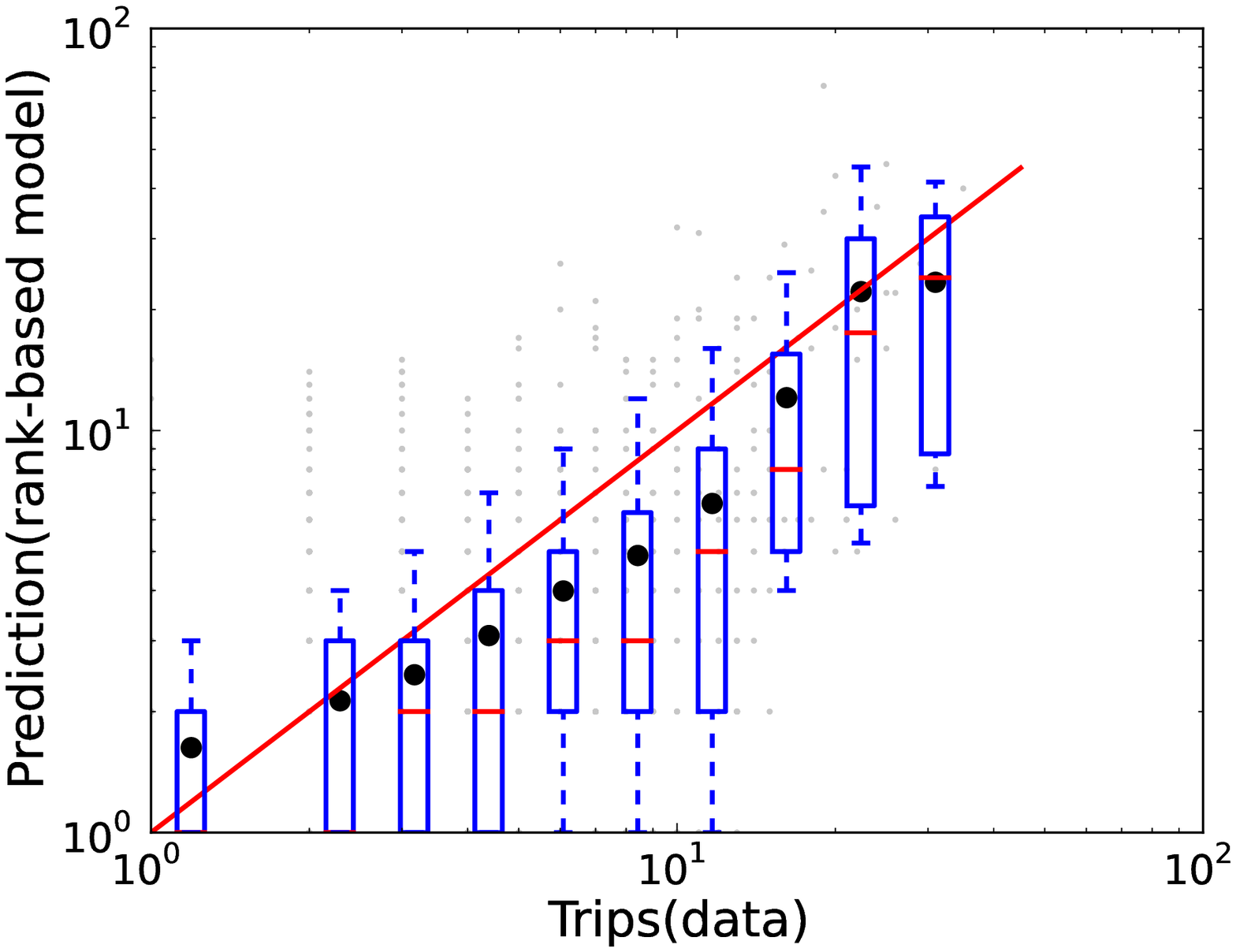}
\includegraphics[scale=.18]{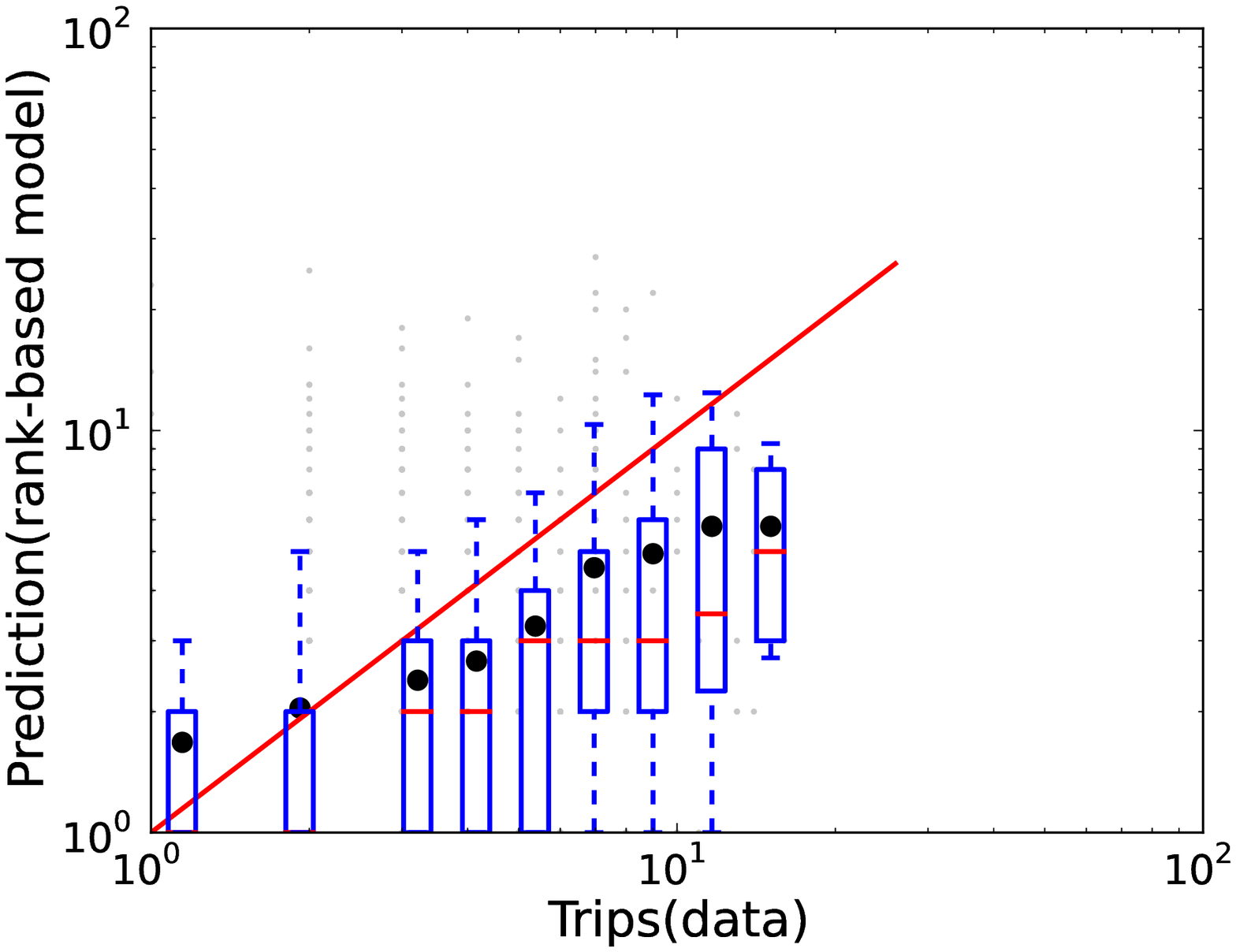}
\includegraphics[scale=.18]{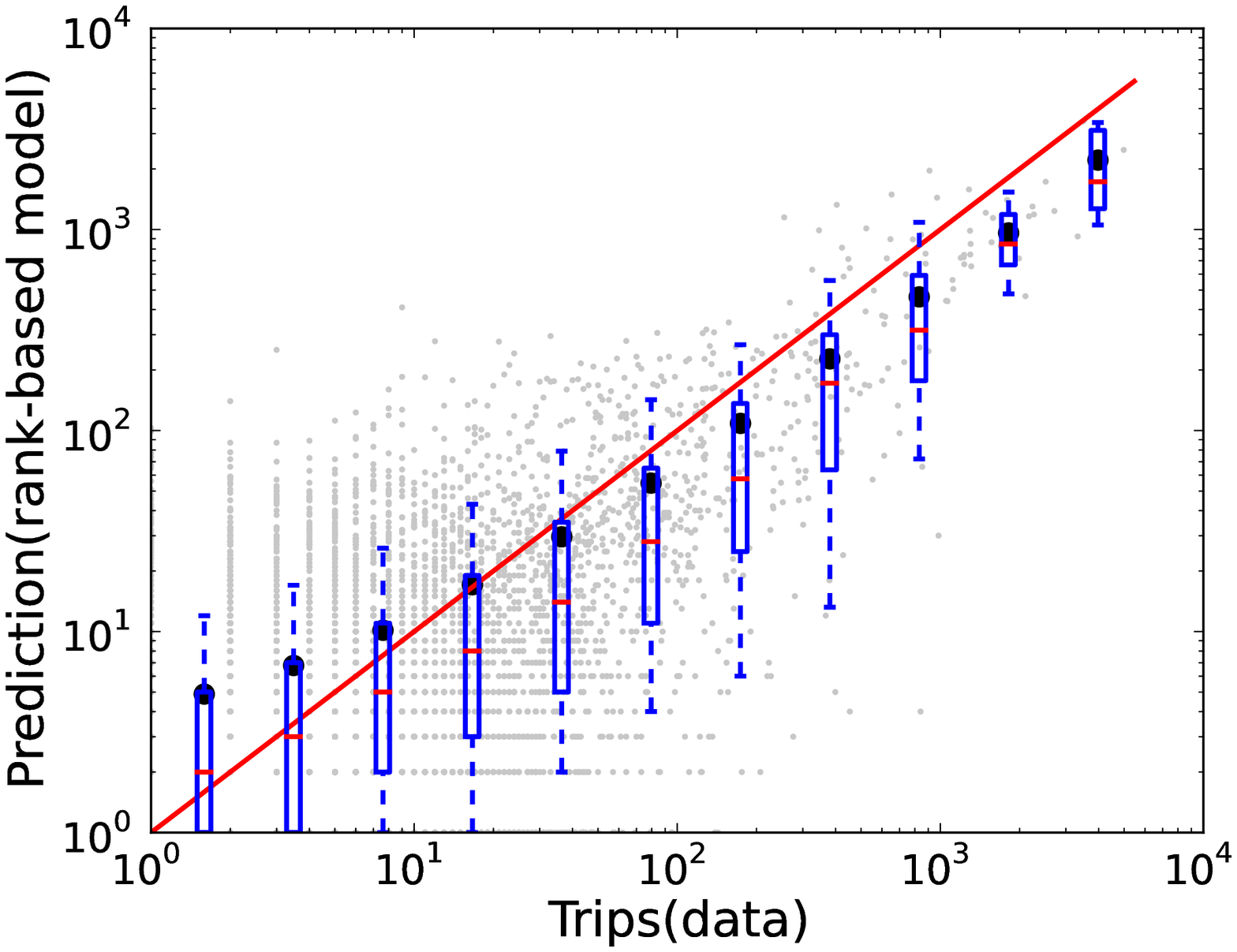}\\
\includegraphics[scale=.18]{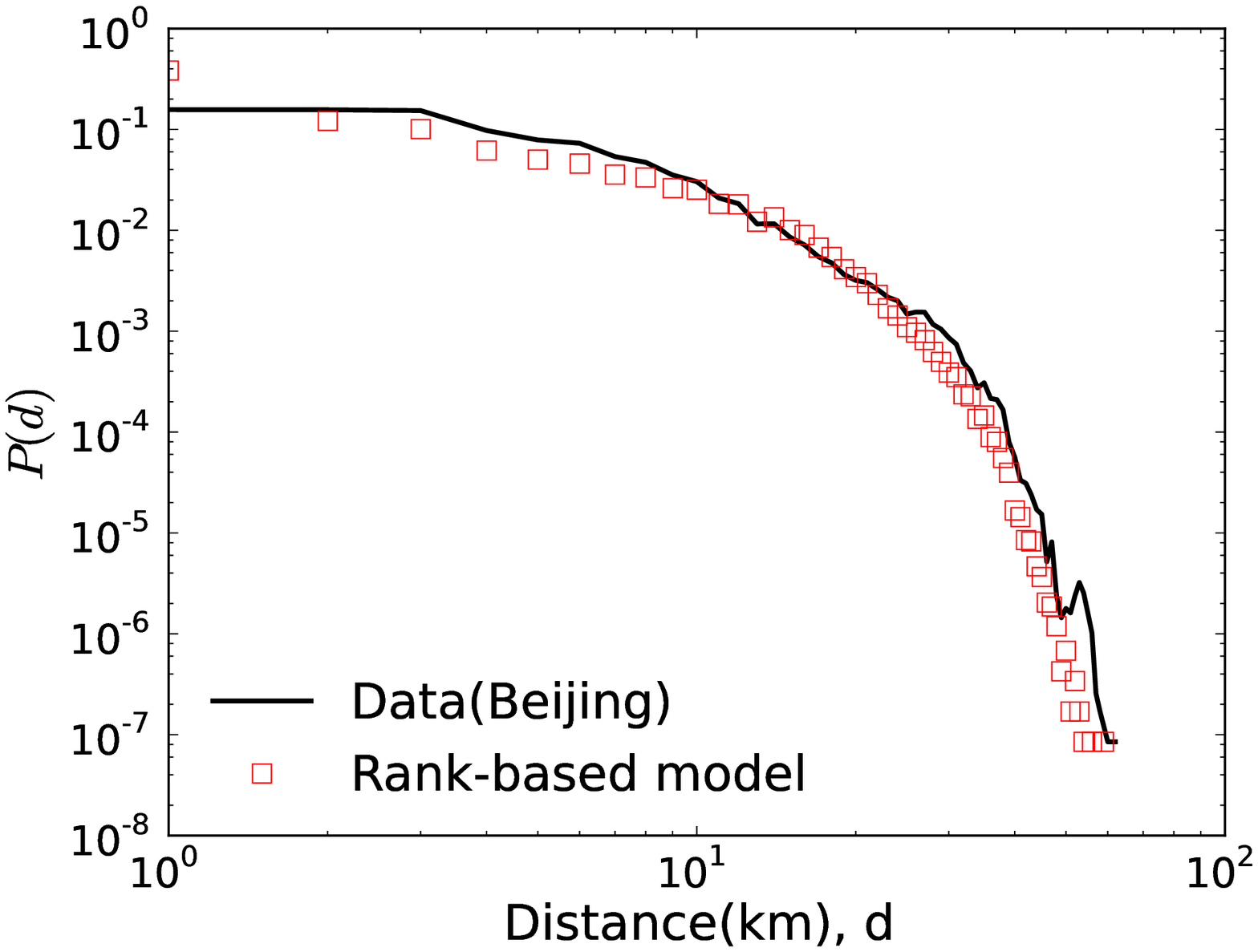}
\includegraphics[scale=.18]{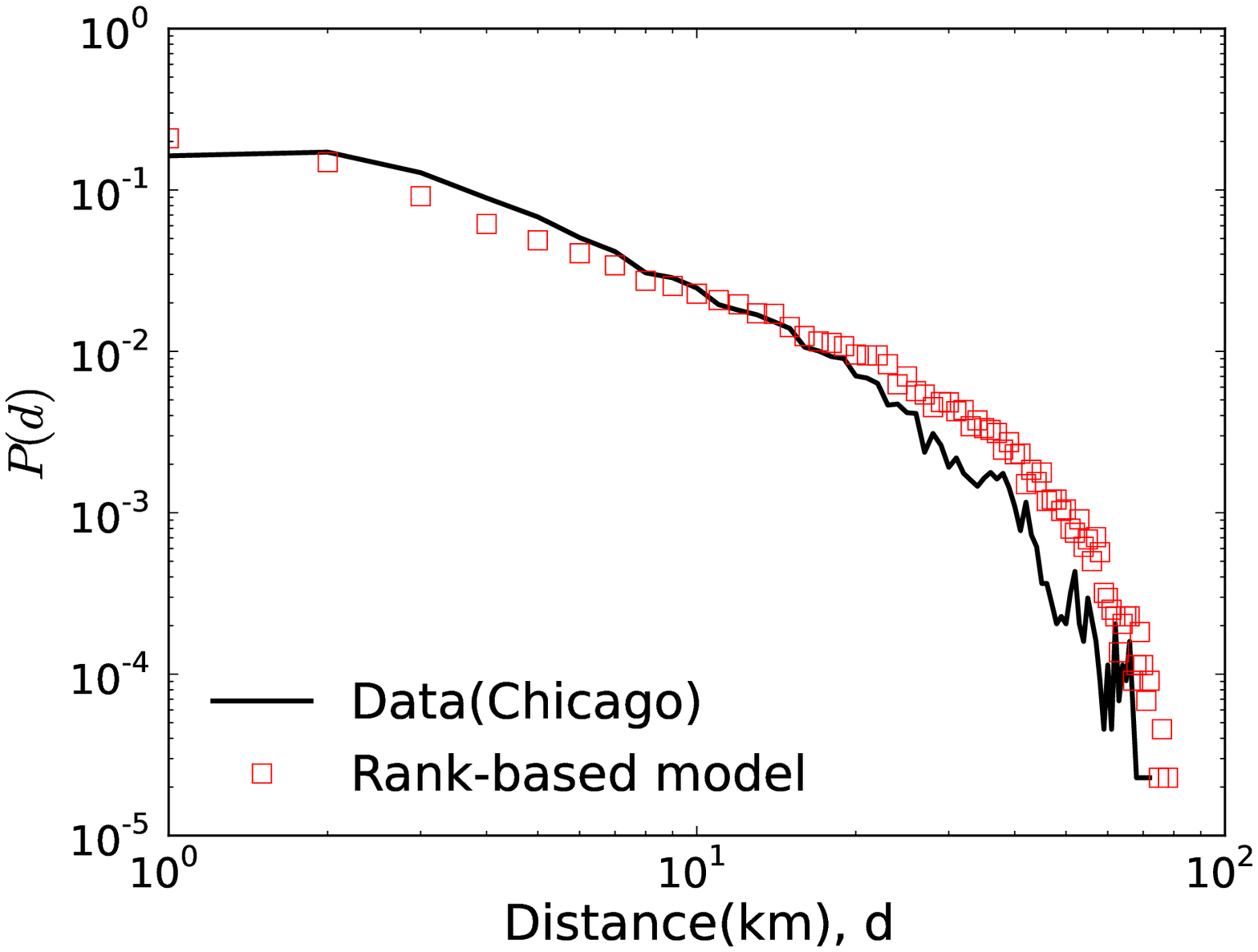}
\includegraphics[scale=.18]{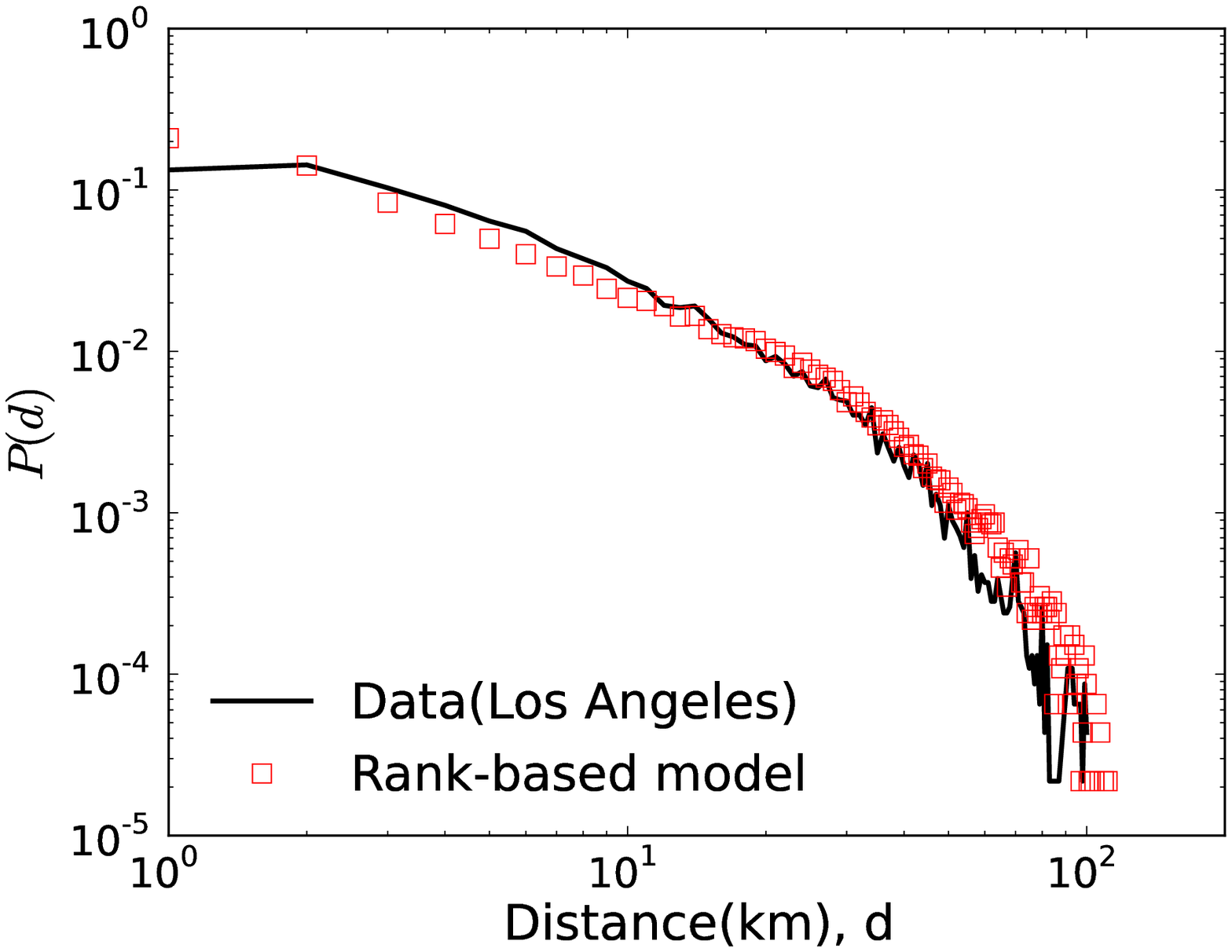}
\includegraphics[scale=.18]{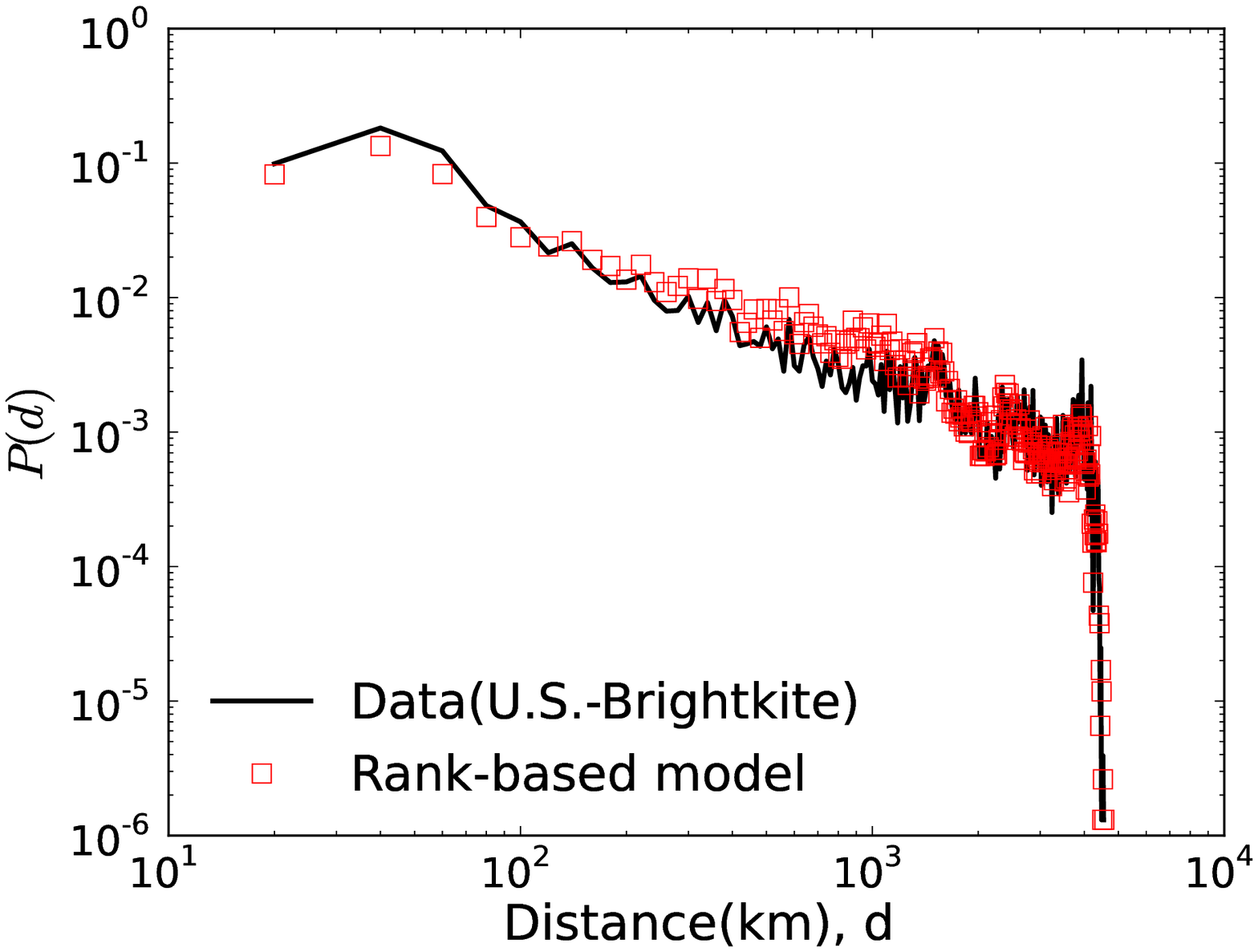}
\caption{{\bf The predicted results by using the rank-based model in the four datasets.} The first row represents the relationships between actual and predicted flows. The second row stands for the comparisons between actual and predicted trip-length distributions. The columns of subgraphs relate to the mobility datasets of Beijing, Chicago, Los Angeles and U.S.(Brightkite) from left to right successively.(The result of U.S. dataset from Gowalla is very similar with the one from Brightkite and thus is not listed.)}
\label{fig:mobility_model}
\end{figure*}

By using Equation (\ref{eqn:prediction}), the predicted results are shown in Fig. \ref{fig:mobility_model}. From the plots in the first row, it can be seen that each red straight line all nearly lies between the whiskers, corresponding to 9th and 91st percentiles in bins, of the box plot. In these datasets, the rank-based model performs slightly worse in Los Angeles than others, which is because of the sparseness of sampled trips between zones. While regarding to the trip-length distribution, as shown in the figure's second row, the actual and forecasted lines are close to each other. In conclusion, it is illustrated that the rank-based model is capable of predicting traffic flows accurately. Meanwhile, the prediction based on the rank-based model only depends on the population of locations and their relative rankings, indicating that collective human mobility is largely driven by the geographical distribution of population. 

\subsection*{Comparisons with other models}
In the previous literature, several different models \cite{Liang2013, Yan} are proposed to model intra-urban human flows between zones. Among them, Liang's model is put forward to explain the origin of exponential law in intra-urban human mobility \cite{Liang2013}. And the conduction model shows its universality in predicting intra-urban mobility patterns and outperforms other models \cite{Yan}. So in the subsection, we mainly compare the rank-based model with the two models. 

As well as the rank-based model, these models can be generalized as a common form: $$T_{ij}=T_i\frac{f(i, j)}{\sum_{k\neq i}{f(i, k)}},$$ where $i, j, k$ represent locations and $f(\cdot, \cdot)$ stands for a function of two locations. For the conduction model \cite{Yan}, $$f(i, j) = P_j(\frac{1}{S_{ji}} - \frac{1}{M}) = P_j(\frac{1}{rank_j(i) + P_j(d_{ji})} - \frac{1}{M}),$$ where $M$ is the total number of population, $d_{ji}$ is the distance between the two locations $i$ and $j$. Meanwhile, $S_{ji}$ is the total number of people in the circle of radius $d_{ji}$ centered at the location $j$ (including the locations $i$ and $j$), $P_j(d_{ji})$ represents the total number of population living $d_{ji}$ away from the location $j$.
For Liang's model \cite{Liang2013}, $$f(i, j) = \frac{P_j}{d(i, j)^\alpha},$$ where $d(i, j)$ is the distance between the locations $i$ and $j$, $\alpha$ is the parameter of the model.
For the rank-based model proposed in this paper, $$f(i, j) = \frac{P_j}{rank_i(j)}.$$

\begin{figure}[htbp]
\centering
\includegraphics[scale=.4]{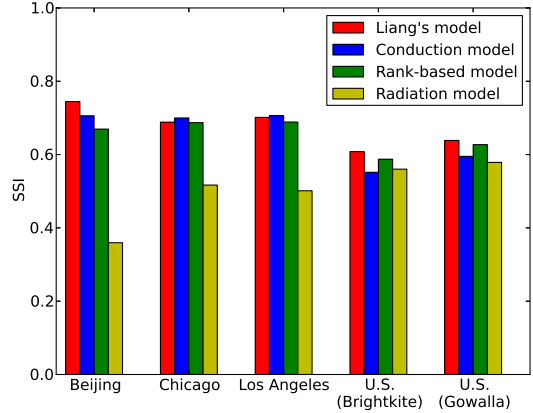}
\caption{{\bf The comparison of different mobility models based on the S\o rensen similarity index.}}
\label{fig:ssi_models}
\end{figure}

So we apply the three models to simulate human mobility for the datasets, the results of prediction and trip-length distributions are shown in Fig. S1 and S2 of Supplementary Information respectively. Although slight differences, these models basically can predict human flows accurately. Here we use the S\o rensen similarity index (SSI) (see details in Materials and Methods) to evaluate these three models as well as the radiation model. As shown in Fig. \ref{fig:ssi_models}, the SSI values of these three models in the datasets almost are larger than 0.6, indicating high prediction accuracy. Moreover, the prediction by the three models is much better than the one by the radiation model, except that the conduction model shows slightly worse precision in the U.S. dataset based on the Brightkite website. Among the three models, Liang's model is often better than the other two models in the datasets except for Chicago and Los Angeles. It is understandable because there is a parameter in Liang's model while the other two models are parameter-free. And the conduction model shows strong ability in predicting human flows when the sample of trips is sparse such as in the travel survey datasets (Chicago and Los Angeles). The conduction model is more focused on the population around the destinations, which results in distinguishing the differences of locations more clearly when the population is sampled sparsely. But the model appears to perform poorly for travels at the country level as the U.S. datasets based on location-based services. Compared with Liang's model and the conduction model, the rank-based model is more stable and only a little worse than Liang's model in these mobility datasets.

Actually, these models could be connected internally through the rank-based law. As we know, population distribution has self-similarity and fractal characteristics \cite{Mandelbrot1982,Appleby1996}. The parameter $\alpha$ of Liang's model is estimated from actual human flows, which is closely relevant to population distribution. We apply the sandbox method \cite{BenAvraham2005} to calculate fractal dimensions of population distributions for these datasets. And the comparison between the values of $\alpha$ and fractal dimensions $D$ is shown in Table \ref{tb:fractal}.
\begin{table}[htbp]
\centering
\caption{\bf{The comparison between the parameter $\alpha$ and fractal dimension $D$.}}
\label{tb:fractal}
\begin{tabular}{cccc}
\hline
Dataset & $\alpha$ & Dimension($D$) \\
\hline
Beijing & 1.60  &  1.52\\
Chicago & 1.83 & 1.38 \\
Los Angeles & 1.81 & 1.58 \\
U.S.(Brightkite) & 1.30 & 1.13\\
U.S.(Gowalla) & 1.21 & 1.08\\
\hline
\end{tabular}
\end{table}
As described in the table, the value of parameter $\alpha$ agrees with the dimension $D$ very well in each dataset except Chicago. The exception may be caused by the problem of data sparseness in the city. Based on the definition of sandbox dimension, $rank_{i}(j)$ can be evaluated by
$rank_{i}(j)\sim d(i, j)^D.$ Therefore, from the rank-based model, the prediction of human flows can be given by
\begin{equation}T_{ij} = T_i\frac{P_j/rank_i(j)}{\sum_{k\neq i}{P_k/rank_i(k)}} \approx T_i\frac{P_j/d(i, j)^D}{\sum_{k\neq i}{P_k/d(i, k)^D}}.\label{eqn:fractal}\end{equation}
The formula has the similar form with Liang's model. So it can account for the reason why fitted value of $\alpha$ is very close to the dimension $D$. This suggests that the parameter $\alpha$ can be determined by the dimension of population distribution in advance without the actual traffic flows. It illustrates that Liang's model is equivalent to the rank-based model.

Furthermore, given the fractal property of population distribution, it can be approximately regarded as $rank_i(j)\approx Cd(i, j)^D \approx rank_j(i)$ for any pair of locations $i$ and $j$. Then we employ it to the conduction model which is given by
\begin{eqnarray*}T_{ij} & = &
T_i\frac{P_j(\frac{1}{rank_j(i)+P_j(d_{ji})}-\frac{1}{M})}{\sum_{k\neq i}{P_k(\frac{1}{rank_k(i)+P_k(d_{ki})}-\frac{1}{M})}}\\
&\approx & T_i\frac{P_j(\frac{1}{Cd(i, j)^D+P_j(d_{ji})}-\frac{1}{M})}{\sum_{k\neq i}P_k(\frac{1}{Cd(i, k)^D+P_k(d_{ki})}-\frac{1}{M})}.
\end{eqnarray*}
It is noticed that $rank_i(k)\approx rank_k(i)\approx Cd(i, k)^D \gg P_k(d_{ki})$ when the location $k$ is far from the location $i$. Meanwhile, M is often a large value. So the estimation of flow $T_{ij}$ is deduced to the following form 
\begin{eqnarray*}T_{ij}
&\approx & T_i\frac{P_j/d(i,j)^D}{\sum_{k\neq i}{P_k/d(i, k)^D}}\\
&\approx & T_i\frac{P_j/rank_i(j)}{\sum_{k\neq i}{P_k/rank_i(k)}},
\end{eqnarray*}
which is exactly the same form as the rank-based model.

Regarding to note the radiation model, which is given by
$$T_{ij} = T_i\frac{P_iP_j}{(P_i+P_{ij})(P_i+P_{ij}+P_j)}.$$
In the formula, $P_{ij}$ is the total population of locations (except $i$ and $j$) which is not farther from the location $i$ than the location $j$. Therefore, the prediction by the radiation model can be approximately written as
\begin{equation}T_{ij} \propto \frac{P_j}{rank_i^2(j)}.\label{eq:radiation}\end{equation}
But, as for the rank-based model, the flux can be described as
$$T_{ij} \propto \frac{P_j}{rank_i(j)}.$$
Actually there is no conflict between these two models and both of them have different application scenarios. From Equation (\ref{eq:radiation}), we can learn that the trips simulated by the radiation model prefer lower ranks, which correspond to shorter distances. And it also explains why the radiation model is more suitable for modeling the flows of commuting or migration (which is more sensitive to the cost determined by the trip length) rather than more frequent movements with various purposes in the daily life.

In summary, when applying Liang's model, the rank-based model and the conduction model to predict human flows, the population distribution are considered implicitly. Just because of this, the three models are equivalent to some extent. In Liang's model, the parameter should be fitted from actual flows or evaluated from the fractal dimension of population distribution in advance. But the calculation of the rank encodes the geographic information of population distribution naturally. Therefore, the rank-based and the conduction models are parameter-free. However, the rank-based model is more direct and simple, which reduces the intensity of calculation a lot.

\subsection*{Mobility vs. friendship}
In fact, the rank-based model enabling to predict human flows accurately is no accident. Liben-Nowell et al. discover the probability of being friends with a certain person is negatively proportional to the number of closer people in the social network consisting of bloggers from Livejournal and prove it is optimal for decentralized search \cite{Liben-Nowell2005}. The similar relation between geography and friendship is also demonstrated in the social network of Facebook \cite{Backstrom}. And the relation indeed exists in the social networks collected from the U.S. datasets based on the location-based services (see Fig. S4 of Supplementary Information). By comparing with the empirical results of human mobility and friendship, we find an interesting phenomenon: the probability of moving to someplace or making friends with anyone living in the same place is in inverse proportion to the rank.

\begin{figure}[htbp]
\centering
\includegraphics[scale=.25]{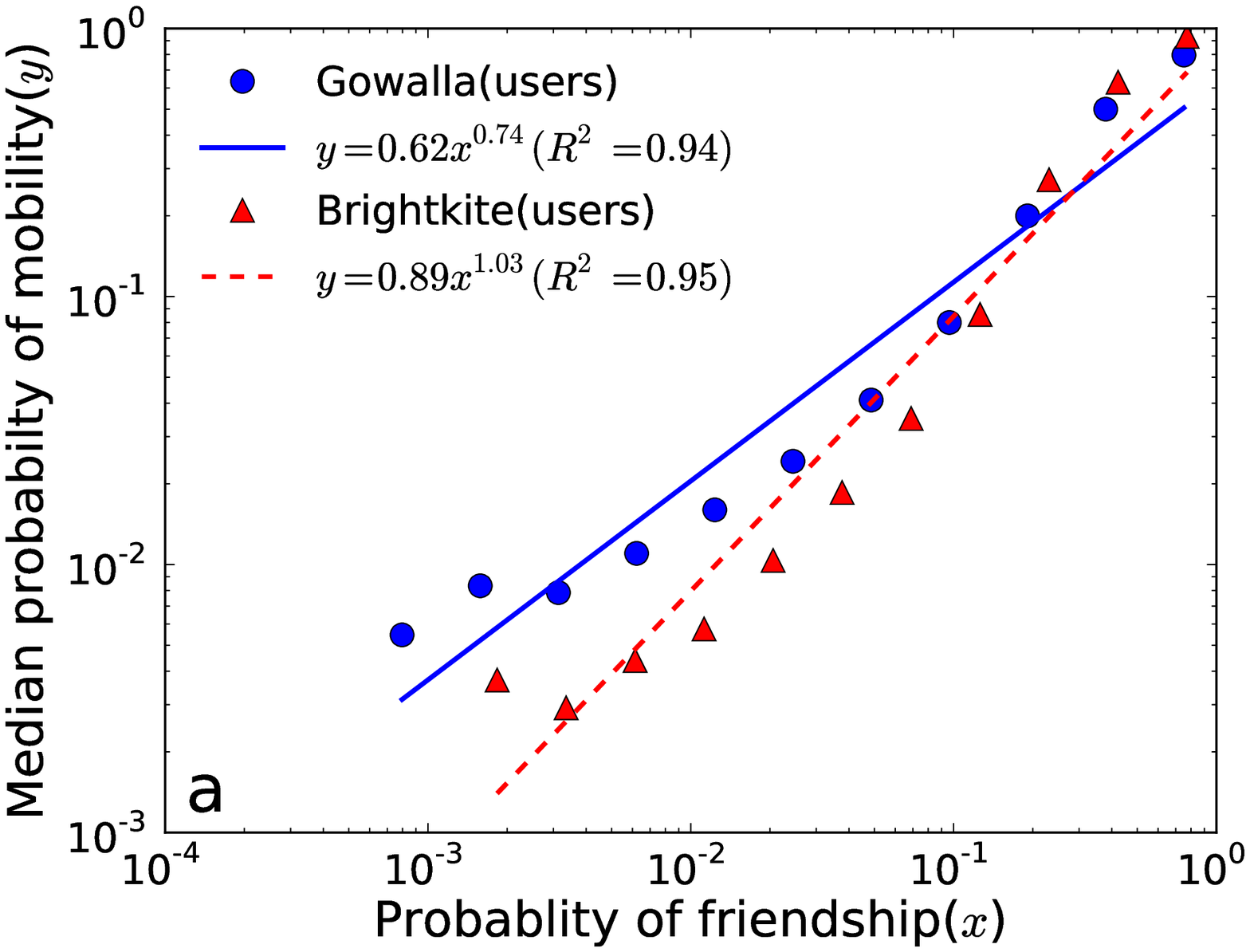}
\includegraphics[scale=.25]{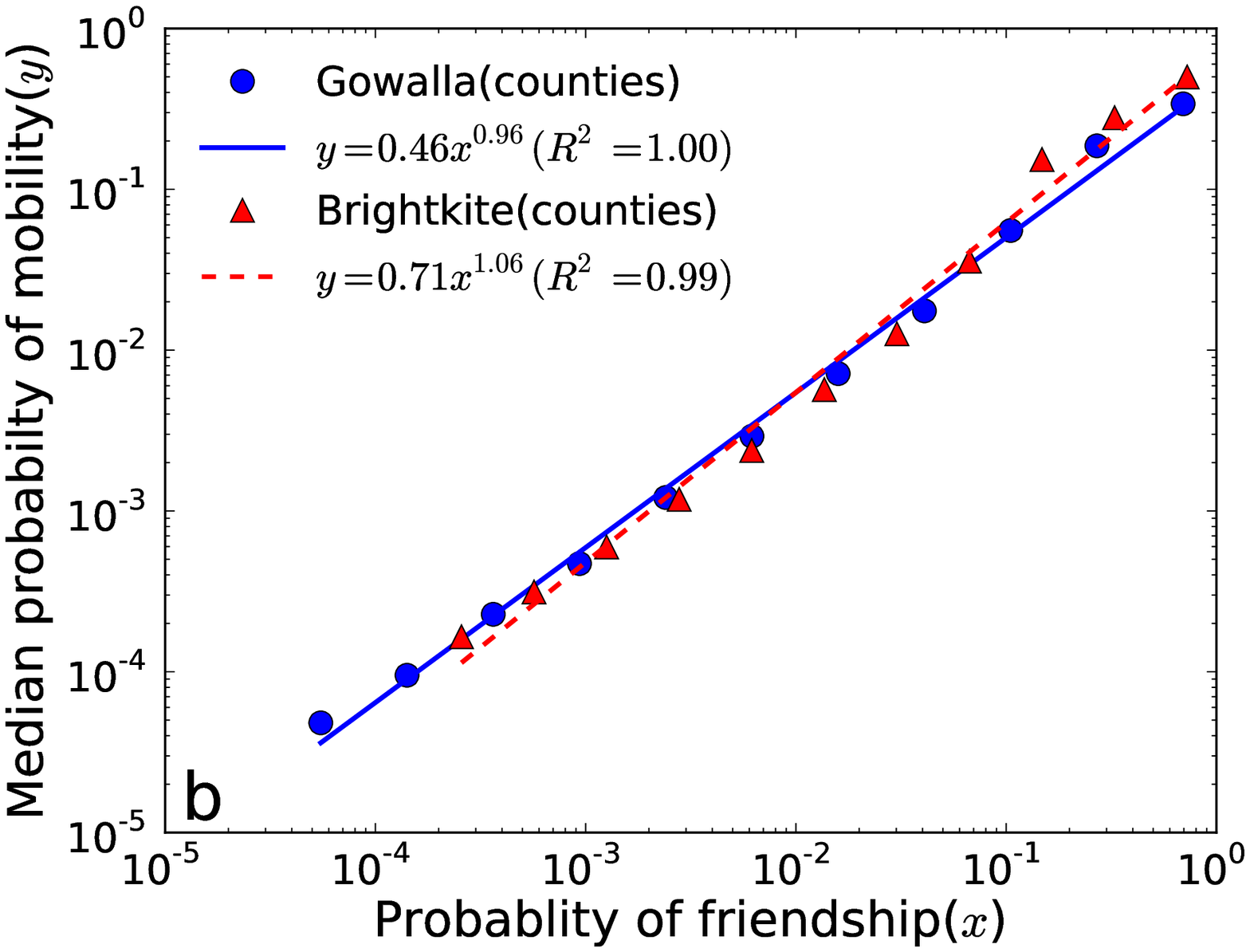}
\caption{{\bf The relationship between friendship and mobility.} (a)At the individual level. (b)At the collective level.}
\label{fig:social_mobility}
\end{figure}

This implies that there may have some subtle connections between human mobility and friendship. Not surprisingly, it seems very reasonable. Due to social purposes, people usually go to visit friends. In turn, when people arrive at unfamiliar locations, they can establish new social relations there. In order to quantify the relationship between friendship and mobility, the fraction of friends in a county and the proportion of trips moving to the same county are compared from two different granularities: each user at the individual level and aggregated all users in each county at the collective level. The results are shown in Fig. \ref{fig:social_mobility}.
From the graphs, it can be seen that the median probabilities of mobility increase with the probabilities of friendship for the two location-based services, revealing the consistent power-law trends with exponents closing to 1.0, especially at the collective level. The strong correlations indicate that the movements have the similar intrinsic mechanism of formation with the friendship. Thus, we believe that there may be some common intrinsic mechanisms to drive the formation of social relationship and mobility. From this perspective, the rank-based model is more fundamental for bridging mobility and friendship perfectly.

\section*{Conclusions}
In this paper, we disclose a universal law in daily human movements, in which the probability from one location to another is inversely proportional to the number of population living in the locations closer than the destination.  Based on this law, a rank-based model is proposed and it can predict human flows accurately at different spatial scales. Out of our expectation, this rule is similar with the regularity of building friendship in social networks. Then we quantify the relationship between mobility and friendship explicitly, which implies the strong correlation between them. Therefore, compared with previous models, the rank-based model is more simple, stable and fundamental. Though the correlation between mobility and friendship is observed, the causality between them is still unclear and deserves more explorations in the future. We also believe that the evolution dynamics of social network is essential in complete understanding how human mobility and social friendship interact with each other.

\section*{Materials and Methods}
\subsection*{Data description}
In this paper, we mainly pay attention to ordinary human movements in cities (counties) or countries. The used datasets captured human travels by taxis, travel surveys and check-ins, whose specific information is introduced as below.
\subsubsection*{Beijing}
In the dataset, the GPS trajectories of passengers were recorded by tracing over 10 thousand taxis in Beijing during three months in 2010.  After dividing the urban areas of Beijing (inside the 6th Ring Road) into squares of 0.01 degree latitude by 0.01 degree longitude, we extracted 11776743 trajectories whose origins and destinations were approximated by the centroids of corresponding squares.
\subsubsection*{Chicago}
The household travel tracker survey data of Chicago from January 2007 to February 2008 are available online at \url{http://www.cmap.illinois.gov/travel-tracker-survey/}. Among them, only the trips within the Cook County, which contains most of the City of Chicago, were considered. Consequently, we obtained a total of 43881 trips between 1314 different census tracts in the county from the survey data. 
\subsubsection*{Los Angeles}
Similar to Chicago, human movement data in Los Angeles come from the household travel survey in 2001 (available online at \url{http://www.scag.ca.gov/travelsurvey/}). From the travel survey, we collected a total of 46000 tracks in the Los Angeles County consisting of 2017 census tracts.

\subsubsection*{United States}
We investigated human mobility in the United States (except Hawaii and Alaska) according to two location-based social networking websites: Brightkite and Gowalla (refer to \url{http://snap.stanford.edu/data/#locnet}).  In both websites, users could build social relationships and share their locations to friends by checking-in services. From the check-in histories of users, we can deduce their home locations approximately (see details in the next subsection). Furthermore, it is noticed that seldom users could share their current positions frequently for protecting privacy and they are more inclined to check into famous or interesting places. Because we emphasize on coarse-grained human travels between counties, it is more reasonable to define a trip as a move from the home of a user to her/his check-in location than between consecutive check-in locations. Respectively, we extracted 764250 and 1220723 trips between 3221 counties from Brightkite and Gowalla.

\subsection*{Inferring the home locations of users}\label{infer_home}
Recently, there are some studies to make inferences about users' homes from their historical check-in records \cite{Cho2011,Noulas2011,Cheng2011}. The main idea is that one user tends to move in the neighborhoods of his/her home more frequently. In this paper, we applied the method in \cite{Cheng2011} to determine the approximated positions of users' homes, which subdivides the possible region iteratively until achieving certain precision. We also inspected the methods in \cite{Cho2011,Noulas2011} and got similar geographic distributions of home locations of users.

\subsection*{The S\o rensen similarity index}\label{ssi_ind}
In this paper, a metric based on the S\o rensen index, which is proposed in Ref. \cite{Lenormand2012}, is utilized to quantify to what extent the predicting model can reproduce the actual traffic flows. The measure is defined as
$$SSI = \frac{2\sum_{i} \sum_{j} {\min(T_{ij}, T^{'}_{ij})}}{\sum_{i} \sum_{j} {T_{ij}} + \sum_{i} \sum_{j} {T^{'}_{ij}}},$$
where $T_{ij}$ and $T^{'}_{ij}$ represent the real and simulated traffic flows from the location $i$ to $j$ separately. Obviously, the measure varies between 0 and 1. The closer the measure is to 1, the more similar the real and predicted fluxes are.

\section*{Acknowledgments}
This research was supported by 863 Program (Grant No. 2012AA011005), SKLSDE (Grant No. SKLSDE-2013ZX-06) and Research Fund for the Doctoral Program of Higher Education of China (Grant No. 20111102110019). XL and JZ thank the Innovation Foundation of BUAA for PhD Graduates (YWF-12-RBYJ-036 and YWF-13-A01-26).



\date{}

\pagestyle{myheadings}

\setcounter{figure}{0}
\makeatletter
\renewcommand{\thefigure}{S\@arabic\c@figure}

\setcounter{table}{0}
\makeatletter
\renewcommand{\thetable}{S\@arabic\c@table}



\begin{flushleft}
{\Large
\textbf{Supplementary information\\A universal law in human mobility}
}
\\
Xiao Liang, 
Jichang Zhao, 
Ke Xu$^{\ast}$
\\
State Key Lab of Software Development Environment, Beihang University, Beijing 100191, P.R. China
\\
$\ast$ E-mail: kexu@nlsde.buaa.edu.cn
\end{flushleft}



\section*{Comparisons of human mobility models}\label{comp_model}
\begin{figure*}[htbp]
\center
\subfloat[Beijing] {
\includegraphics[scale=.22]{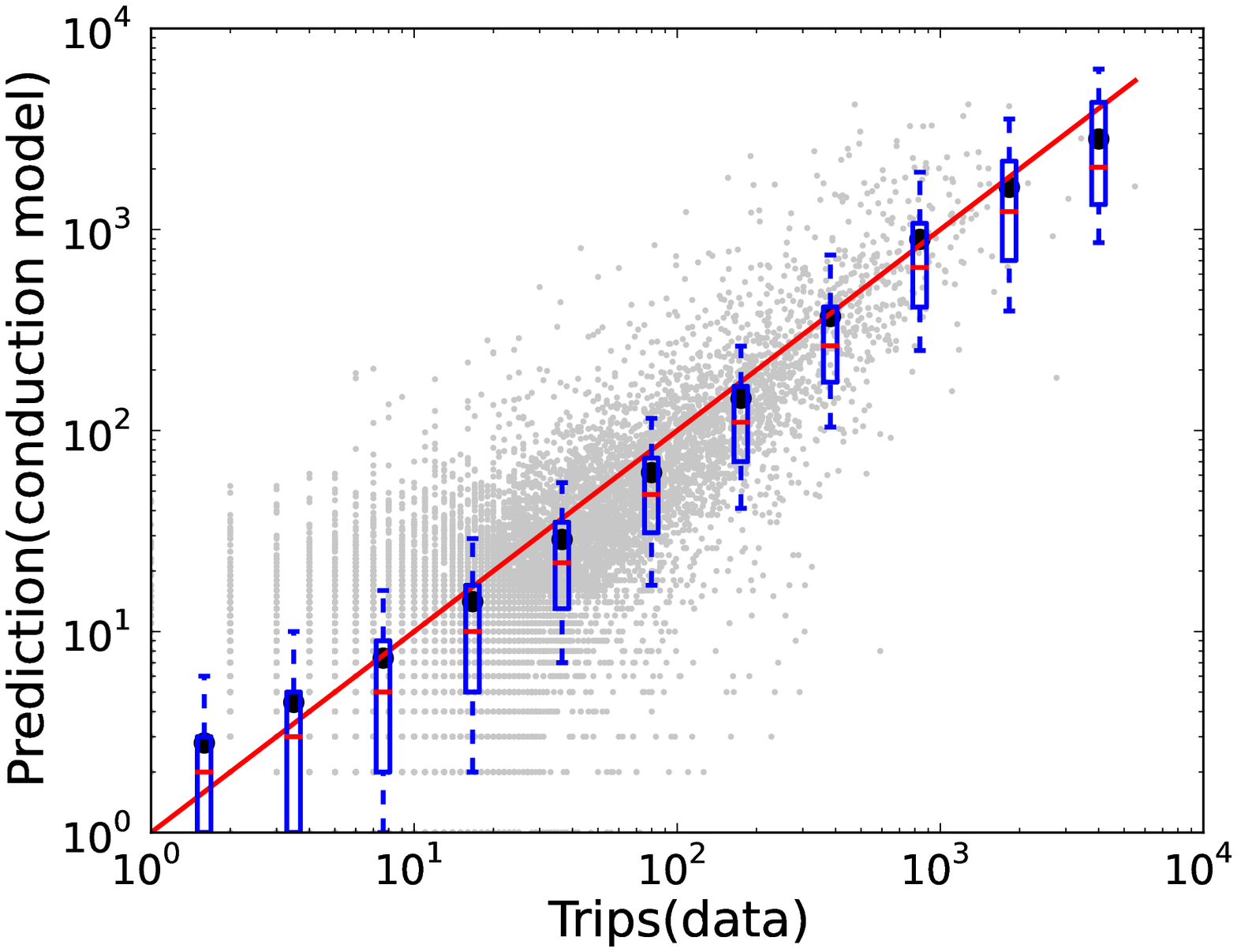}
\includegraphics[scale=.22]{sim_rank_beijing_comp.eps}
\includegraphics[scale=.22]{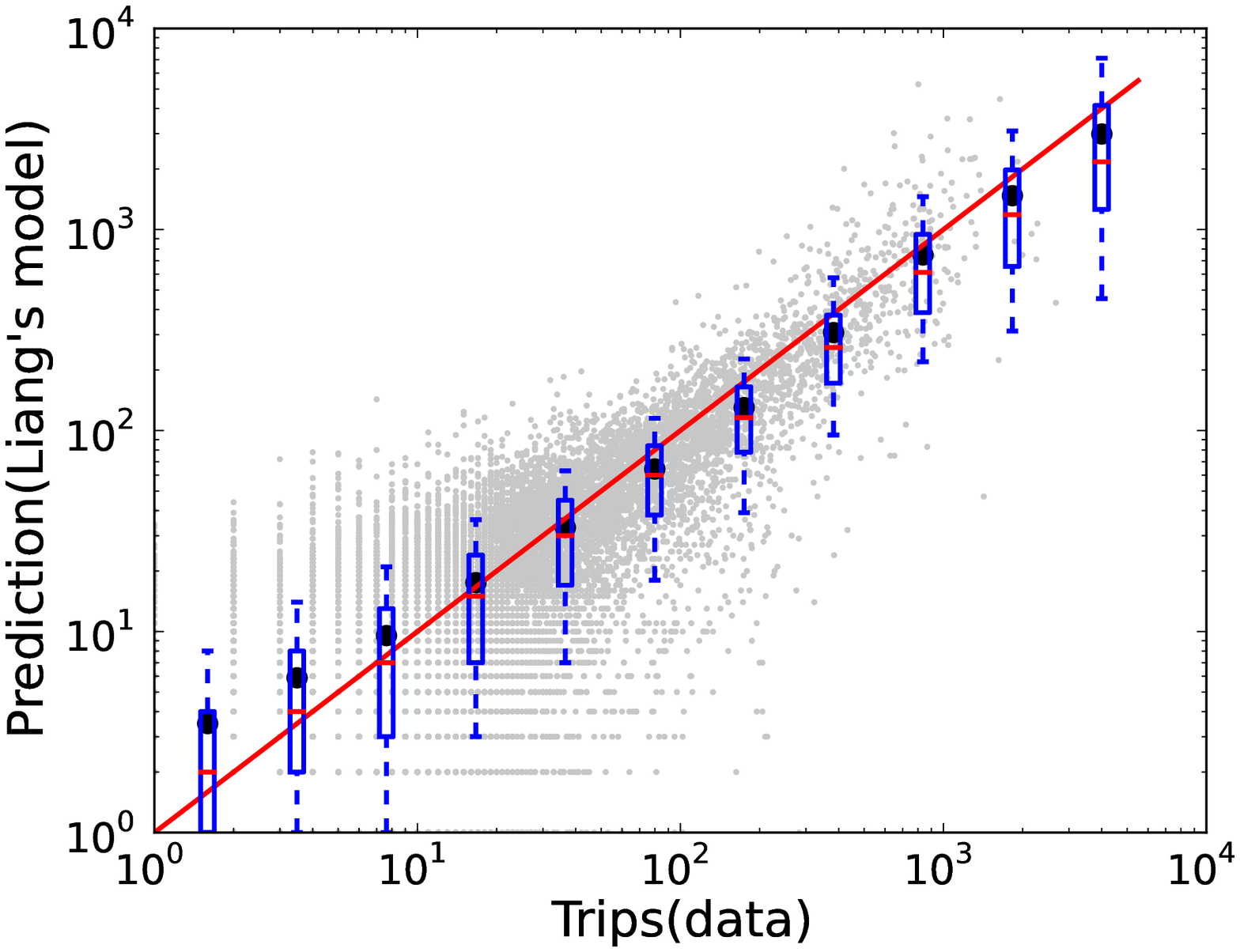}
}\\
\subfloat[Chicago] {
\includegraphics[scale=.22]{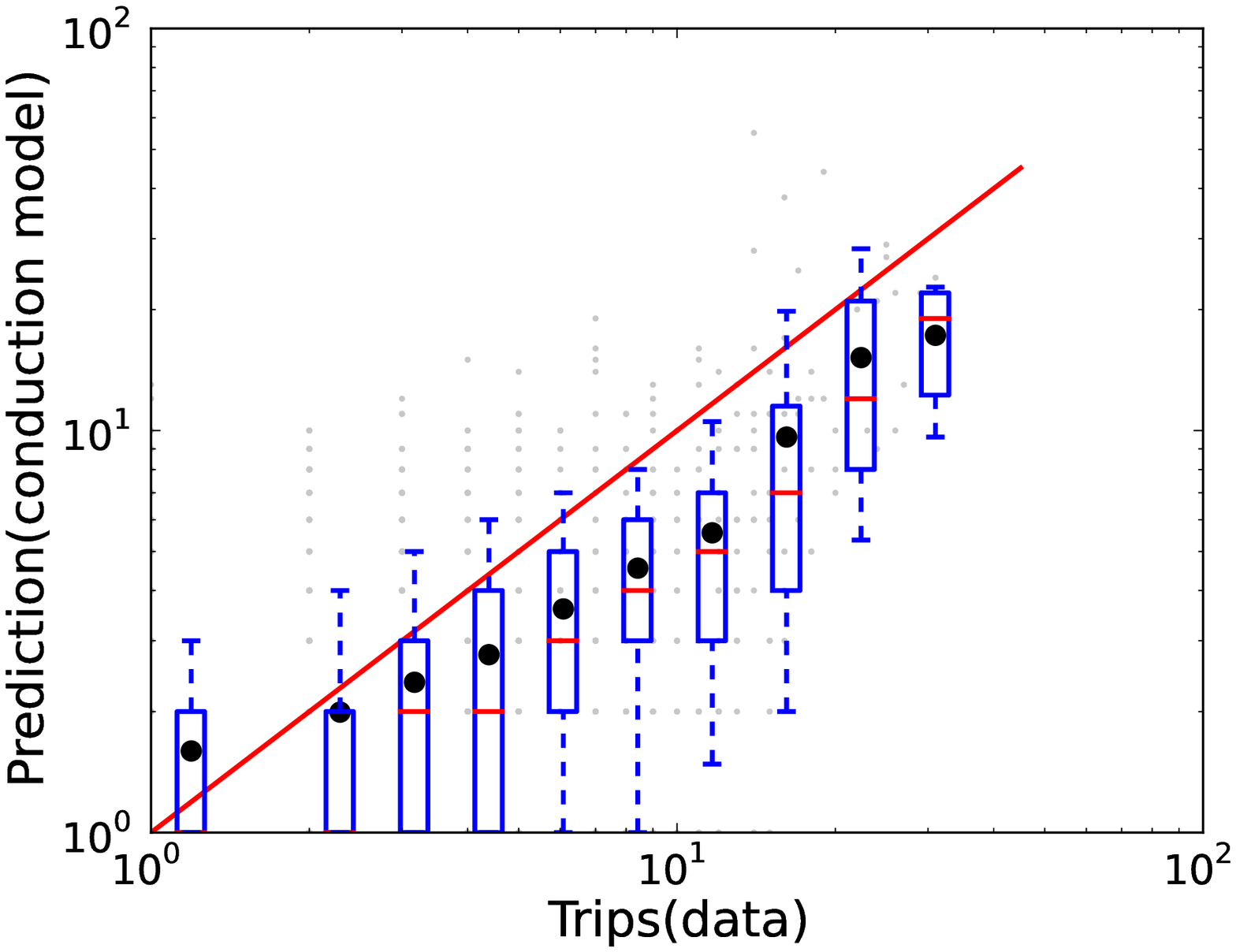}
\includegraphics[scale=.22]{sim_rank_chicago_comp.eps}
\includegraphics[scale=.22]{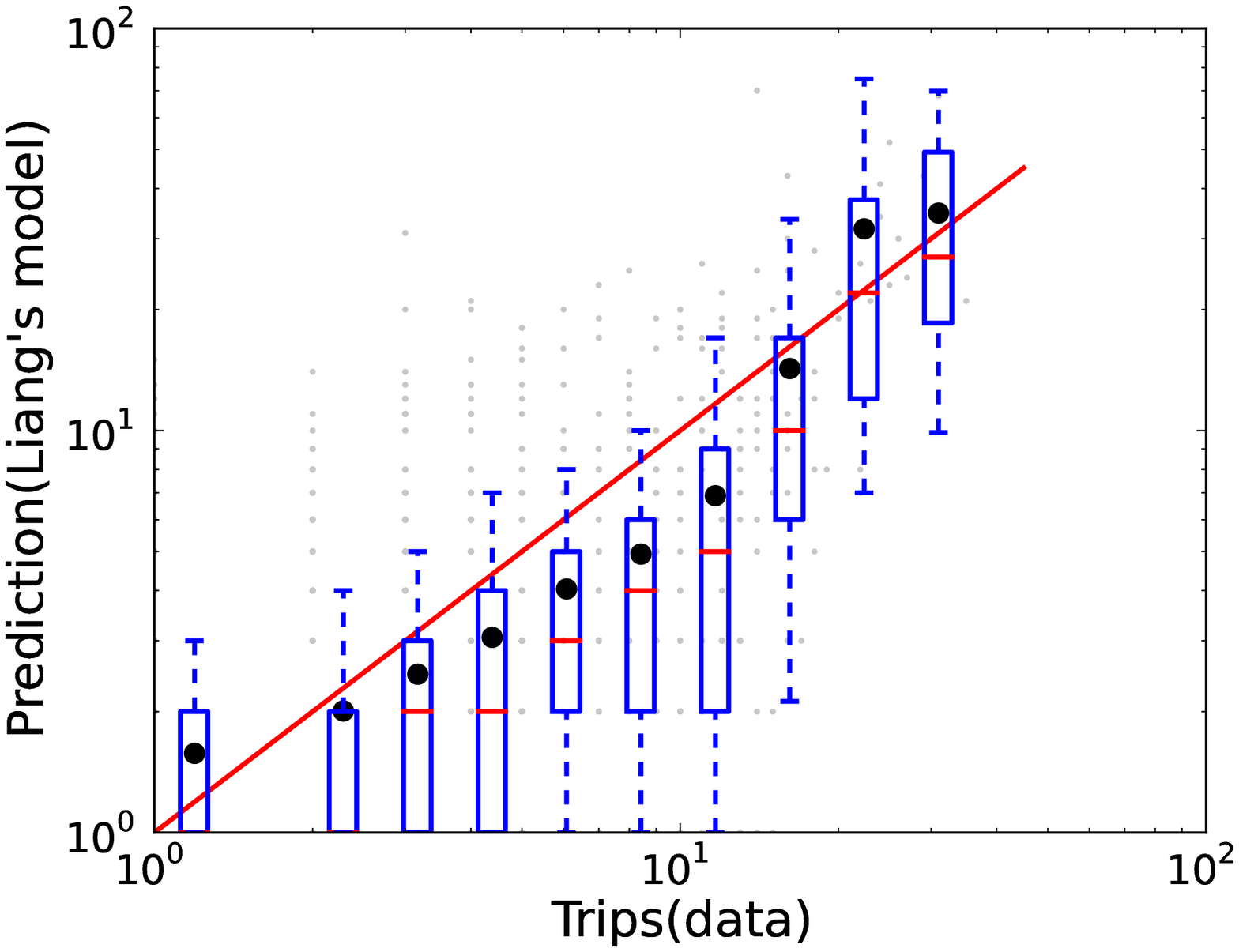}
}\\
\subfloat[Los Angeles] {
\includegraphics[scale=.22]{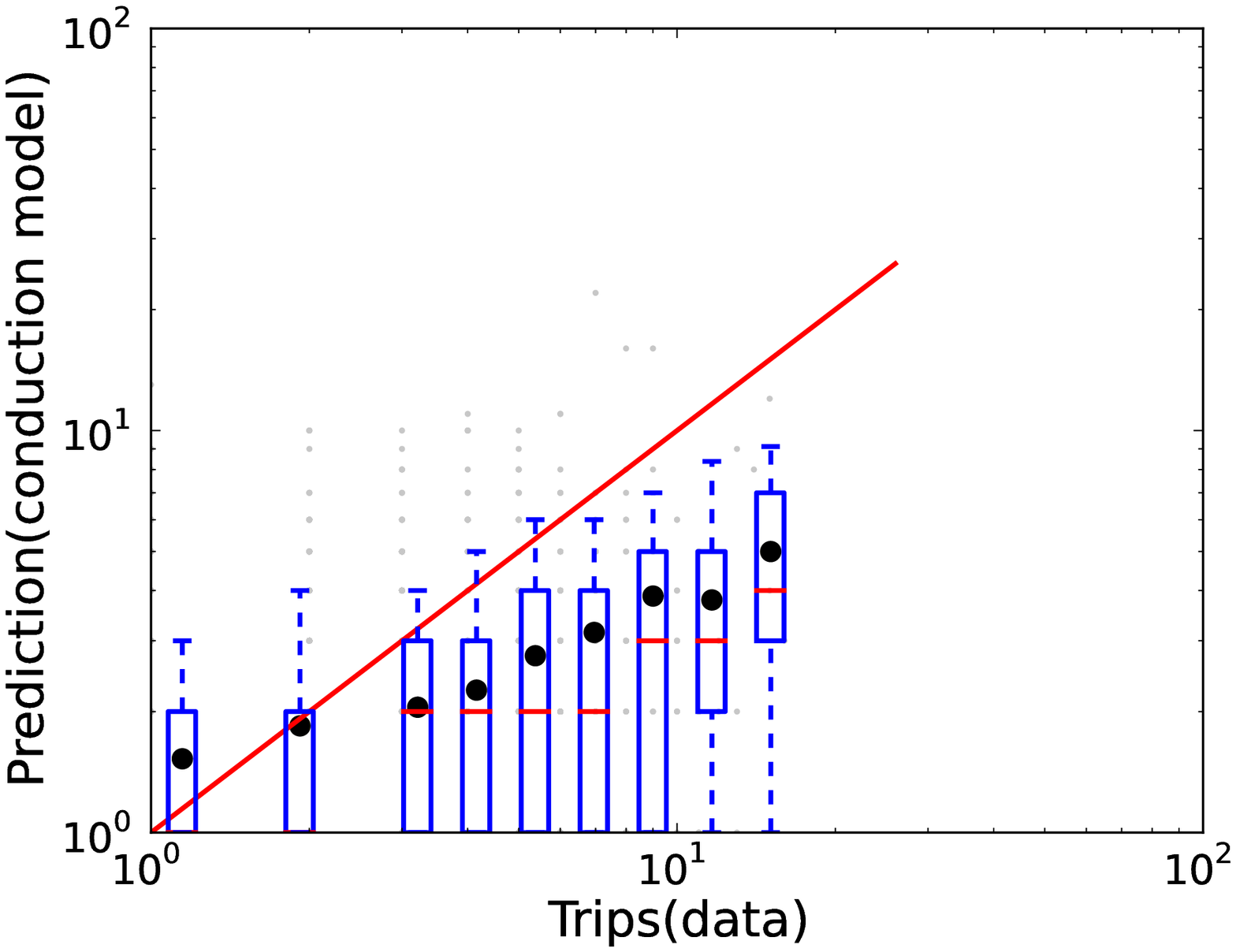}
\includegraphics[scale=.22]{sim_rank_losangeles_comp.eps}
\includegraphics[scale=.22]{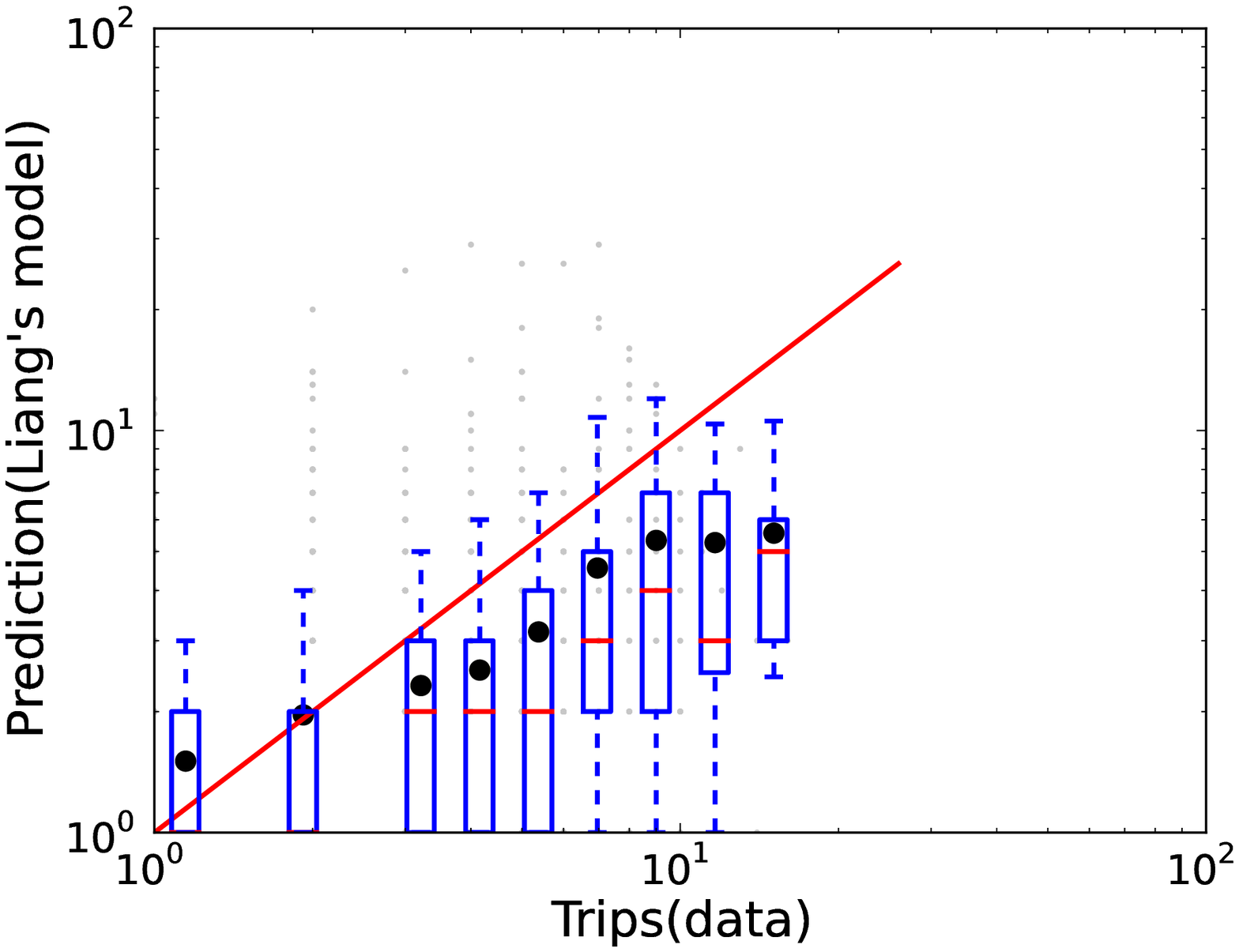}
}\\
\subfloat[U.S.(Brightkite)] {
\includegraphics[scale=.22]{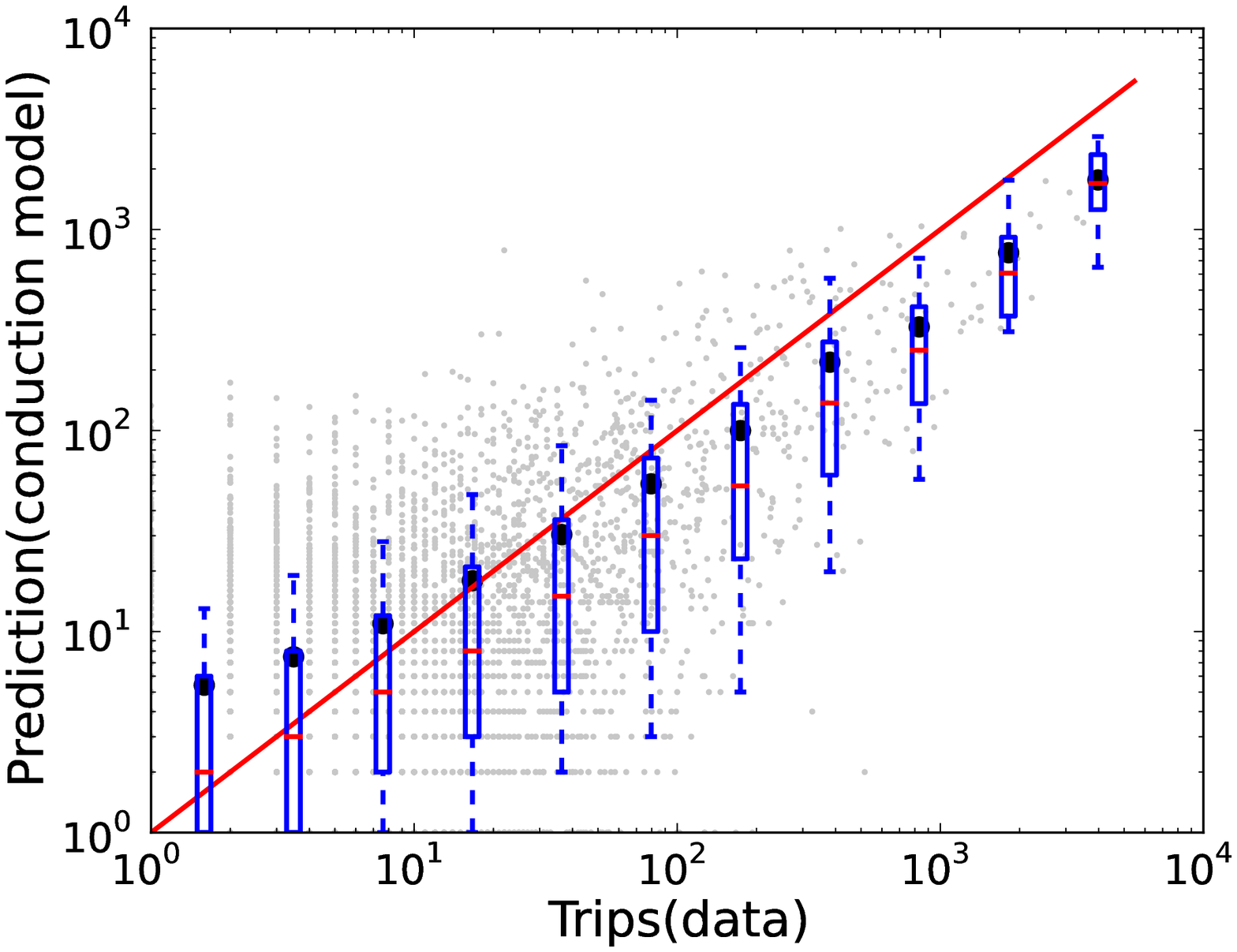}
\includegraphics[scale=.22]{sim_rank_brightkite_us_comp.eps}
\includegraphics[scale=.22]{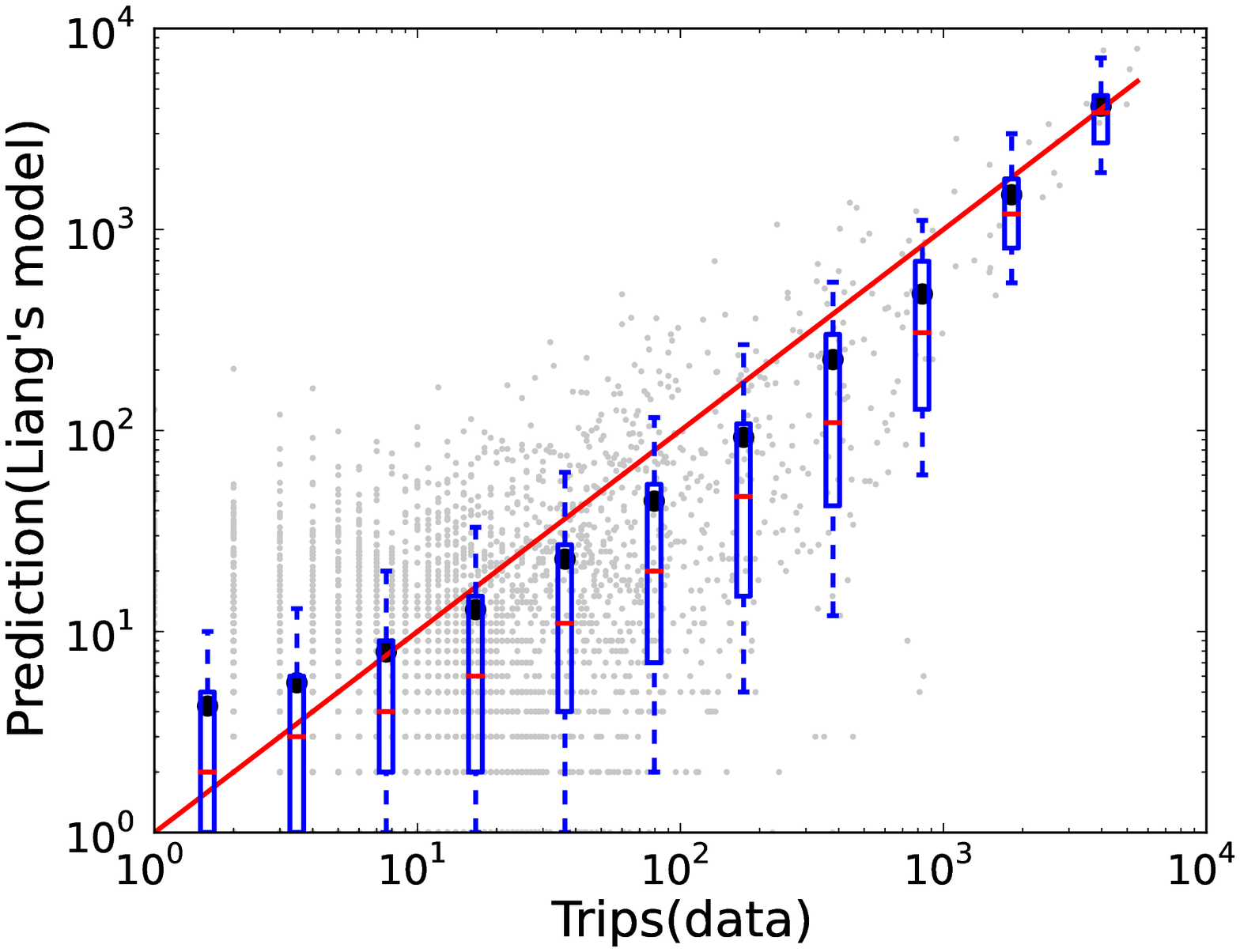}
}\\
\subfloat[U.S.(Gowalla)] {
\includegraphics[scale=.22]{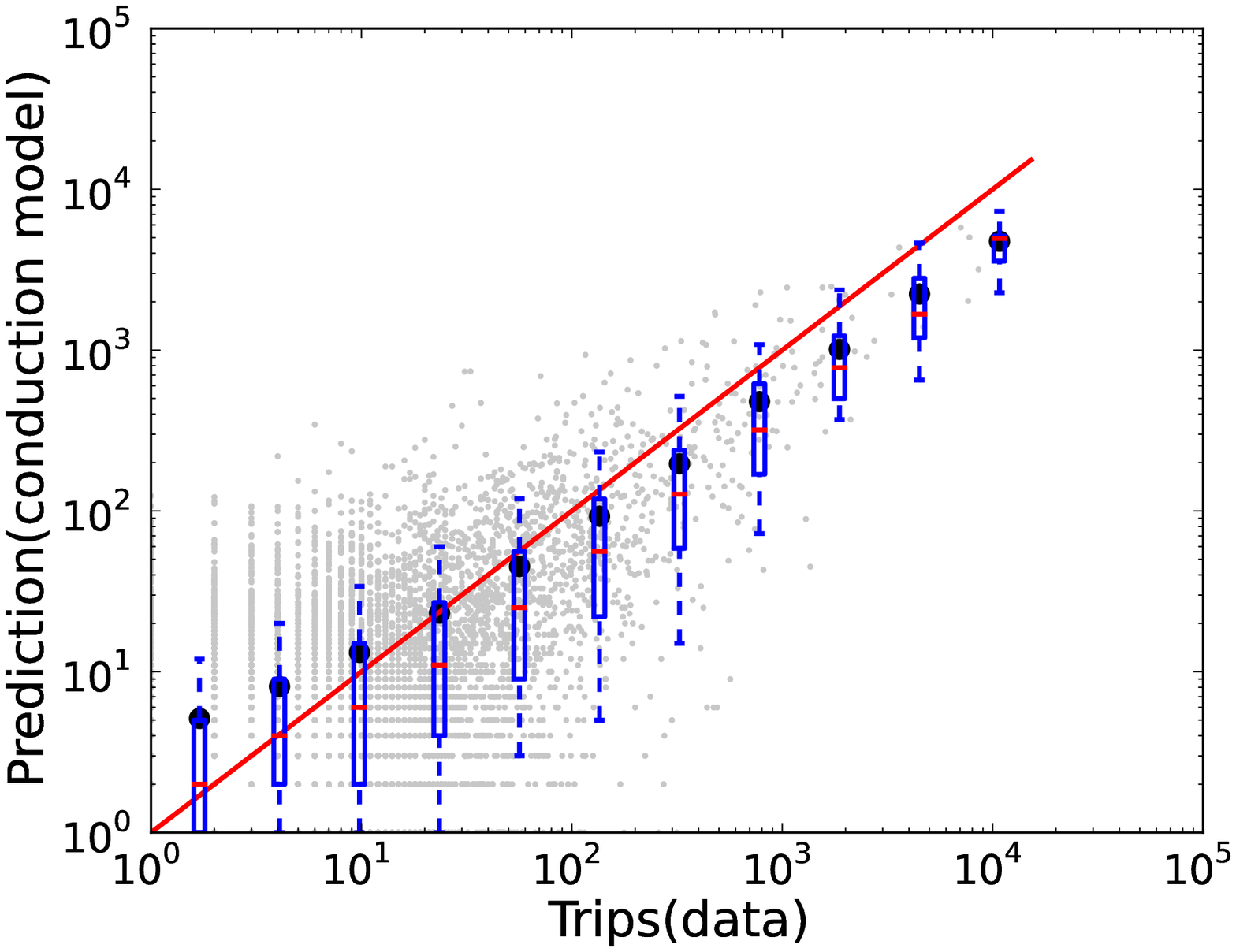}
\includegraphics[scale=.22]{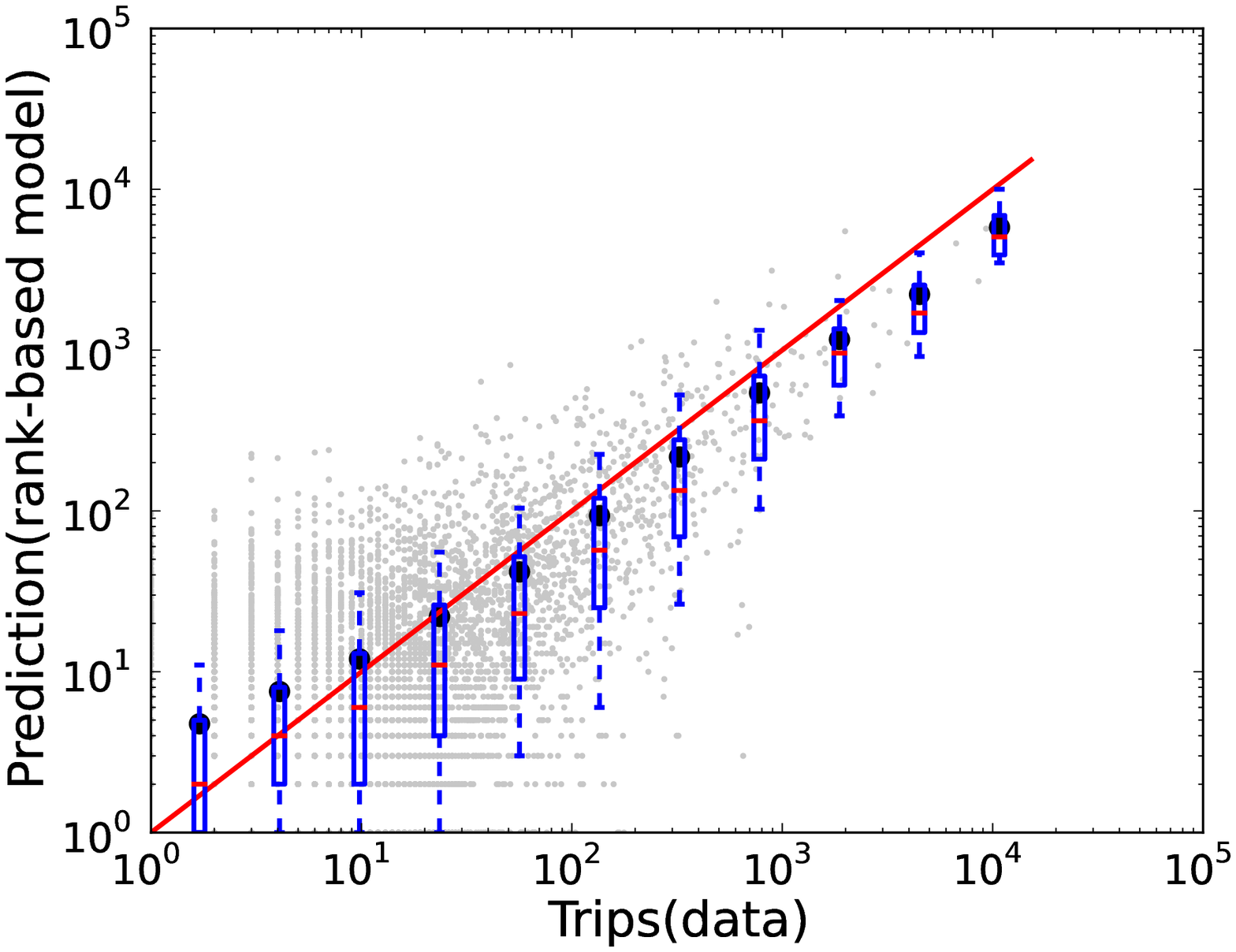}
\includegraphics[scale=.22]{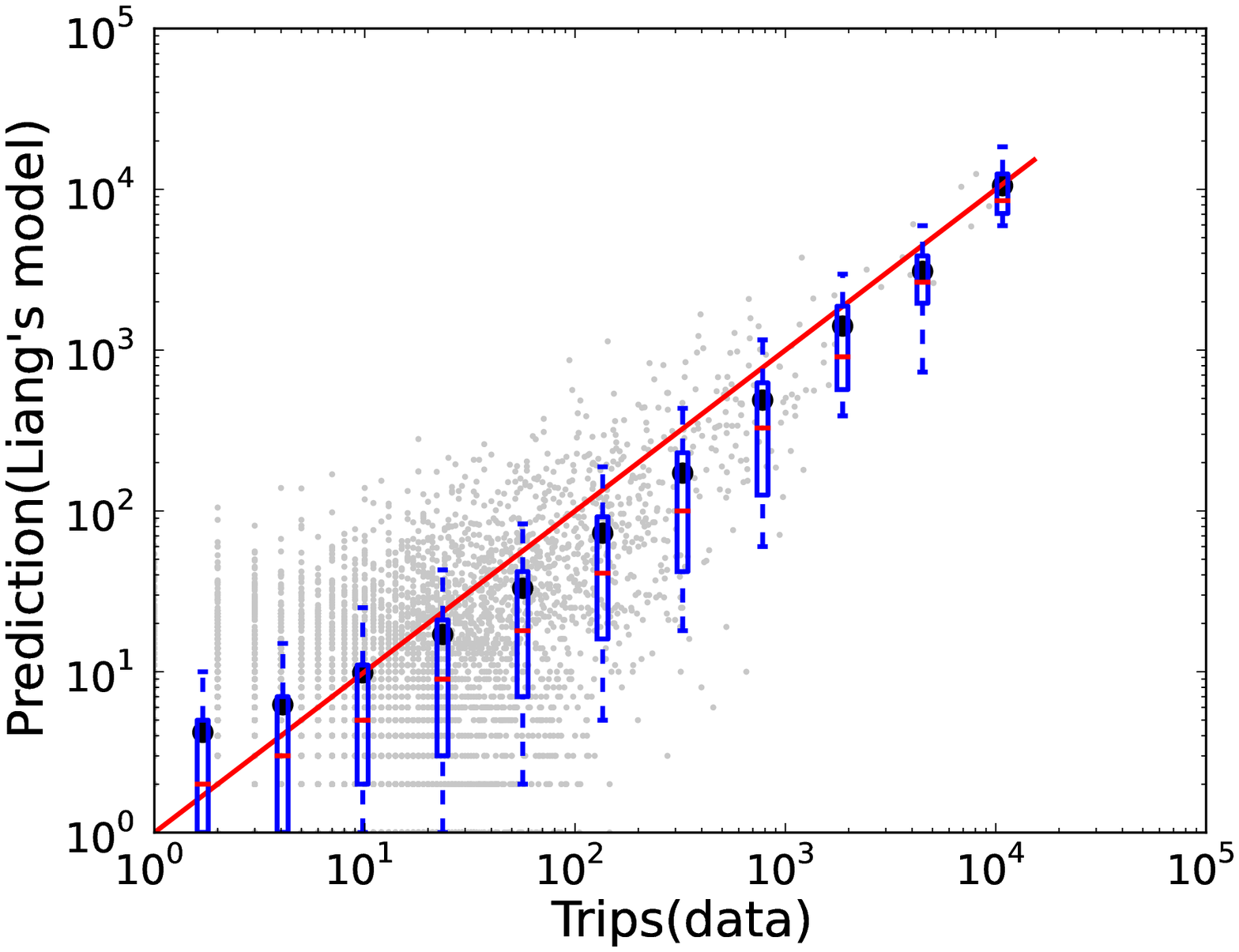}
}
\caption{{\bf The prediction of human flows.} (a)Beijing. (b)Chicago. (c)Los Angeles. (d)U.S.(Brightkite). (e)U.S.(Gowalla). For each dataset, the results of prediction based on the three models-conduction model, rank-based model and Liang's model-are shown in turn.}
\label{fig:flow_comp}
\end{figure*}

\begin{figure*}[htbp]
\subfloat[Beijing] {
\includegraphics[scale=.25]{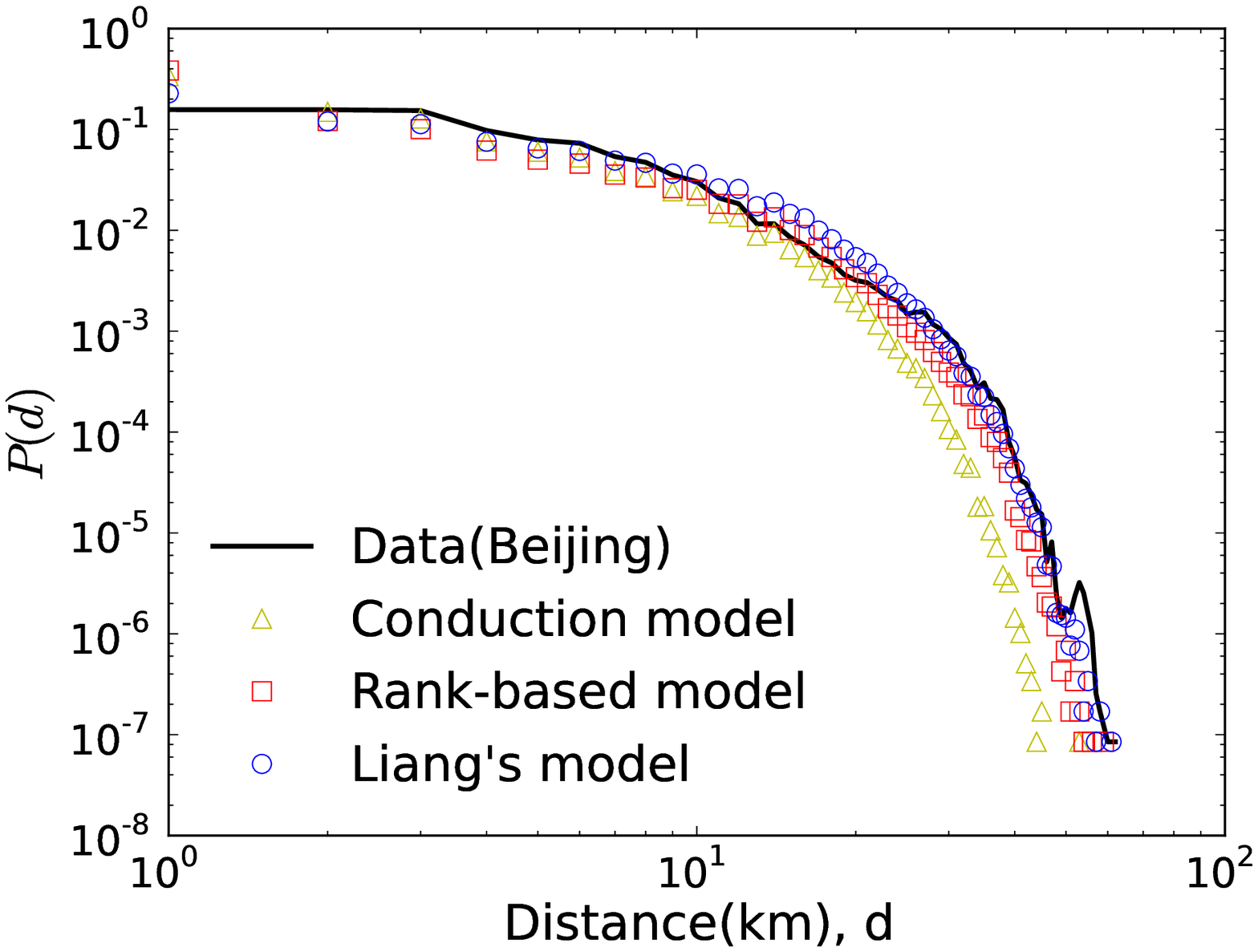}
}
\subfloat[Chicago] {
\includegraphics[scale=.25]{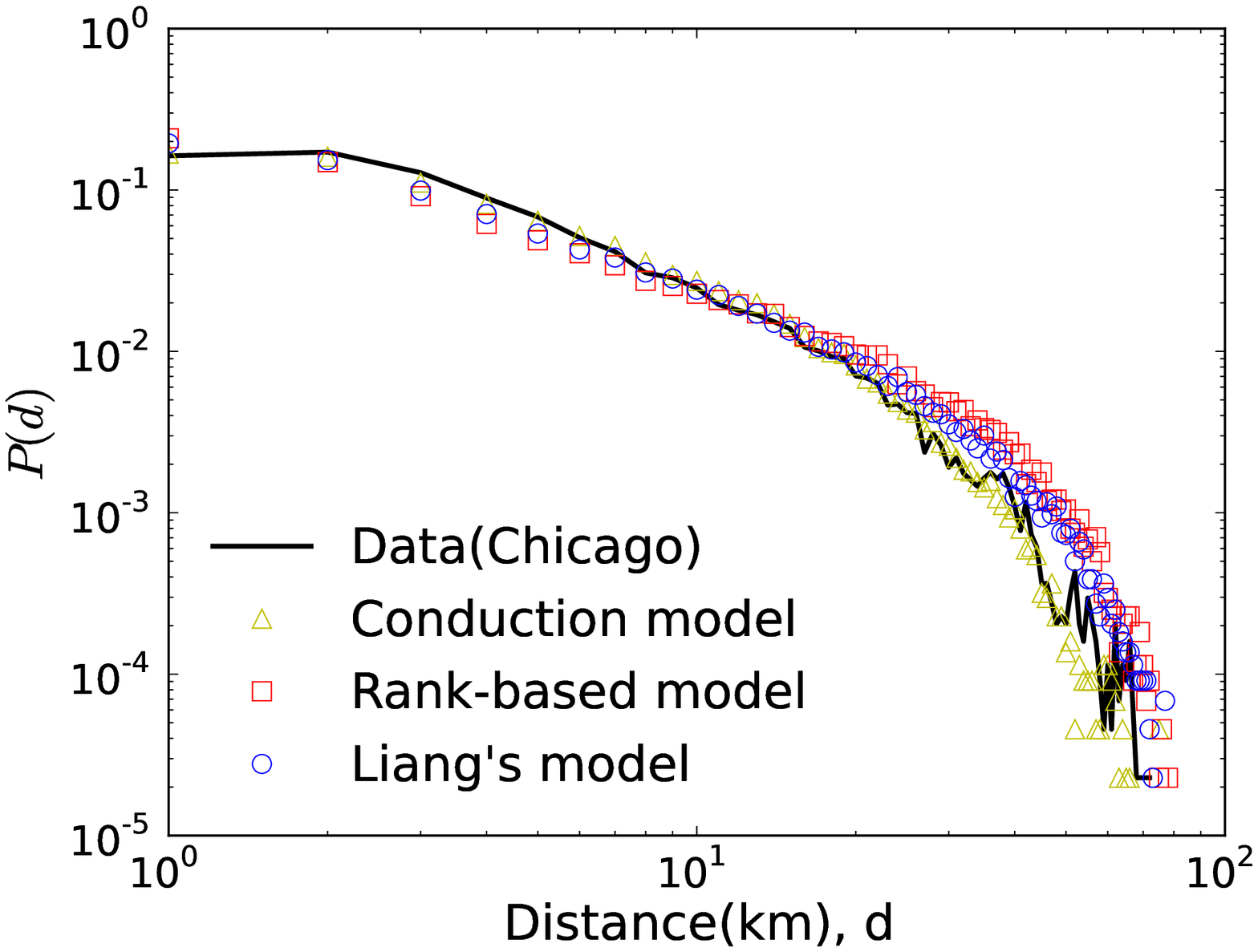}
}
\subfloat[Los Angeles] {
\includegraphics[scale=.25]{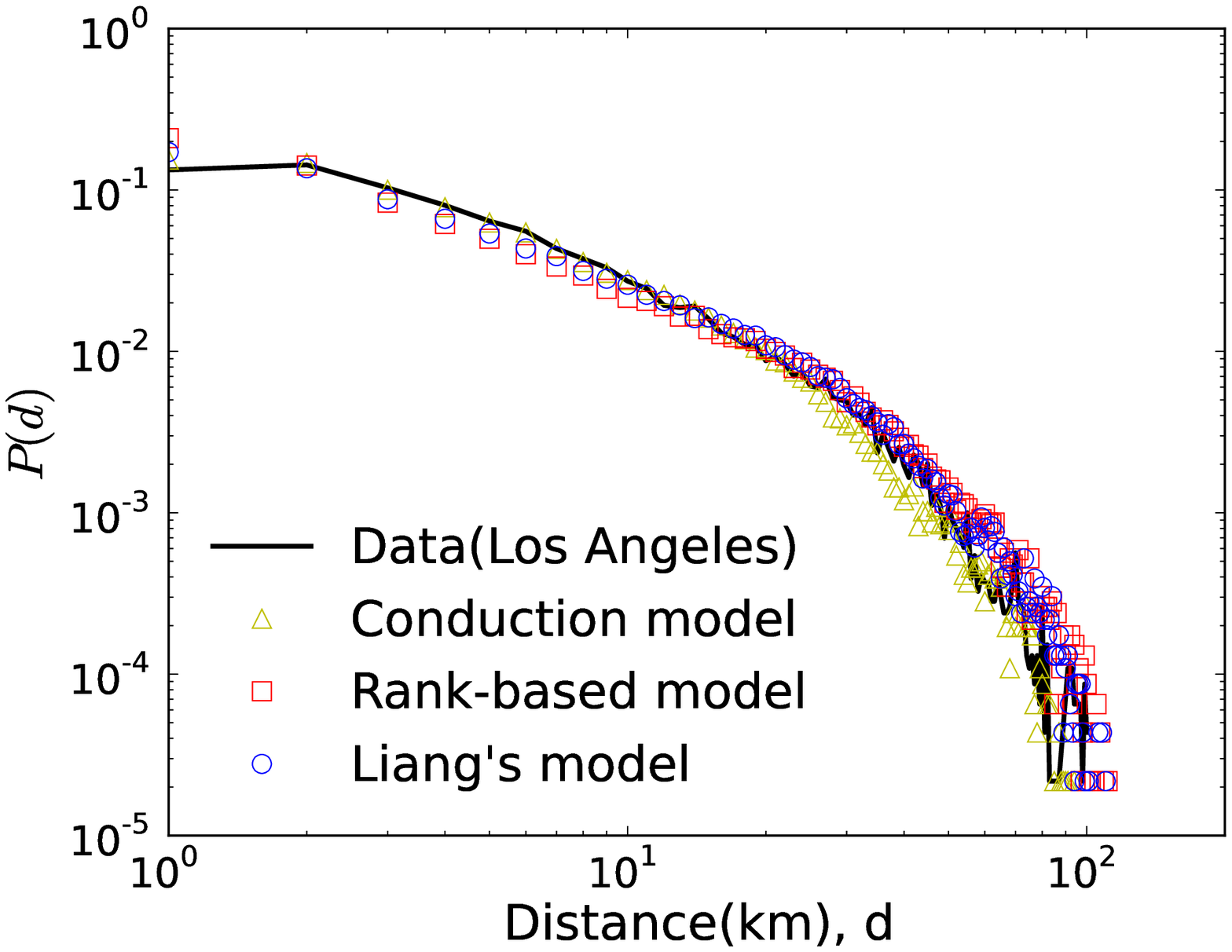}
}\\
\subfloat[U.S.(Brightkite)] {
\includegraphics[scale=.25]{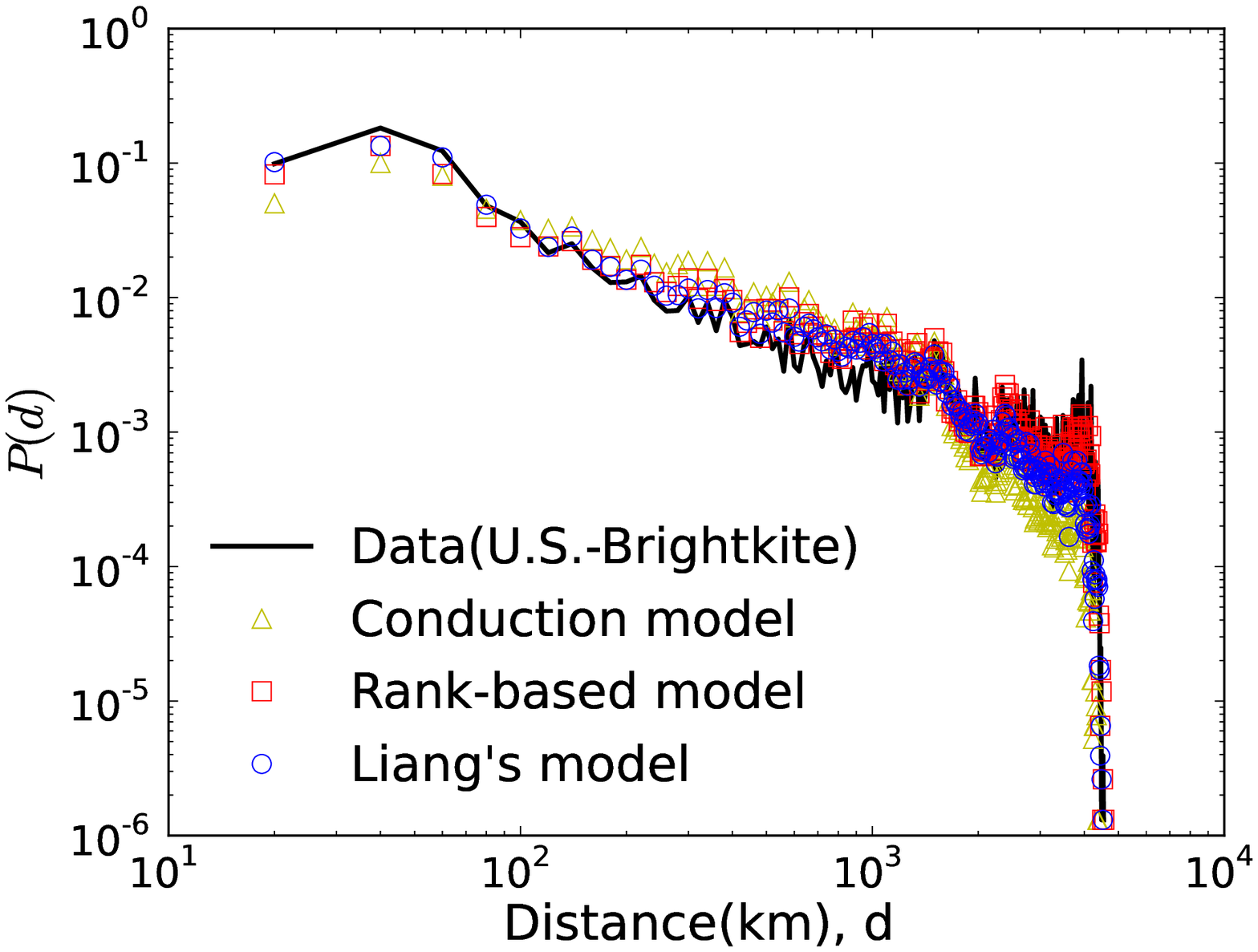}
}
\subfloat[U.S.(Gowalla)] {
\includegraphics[scale=.25]{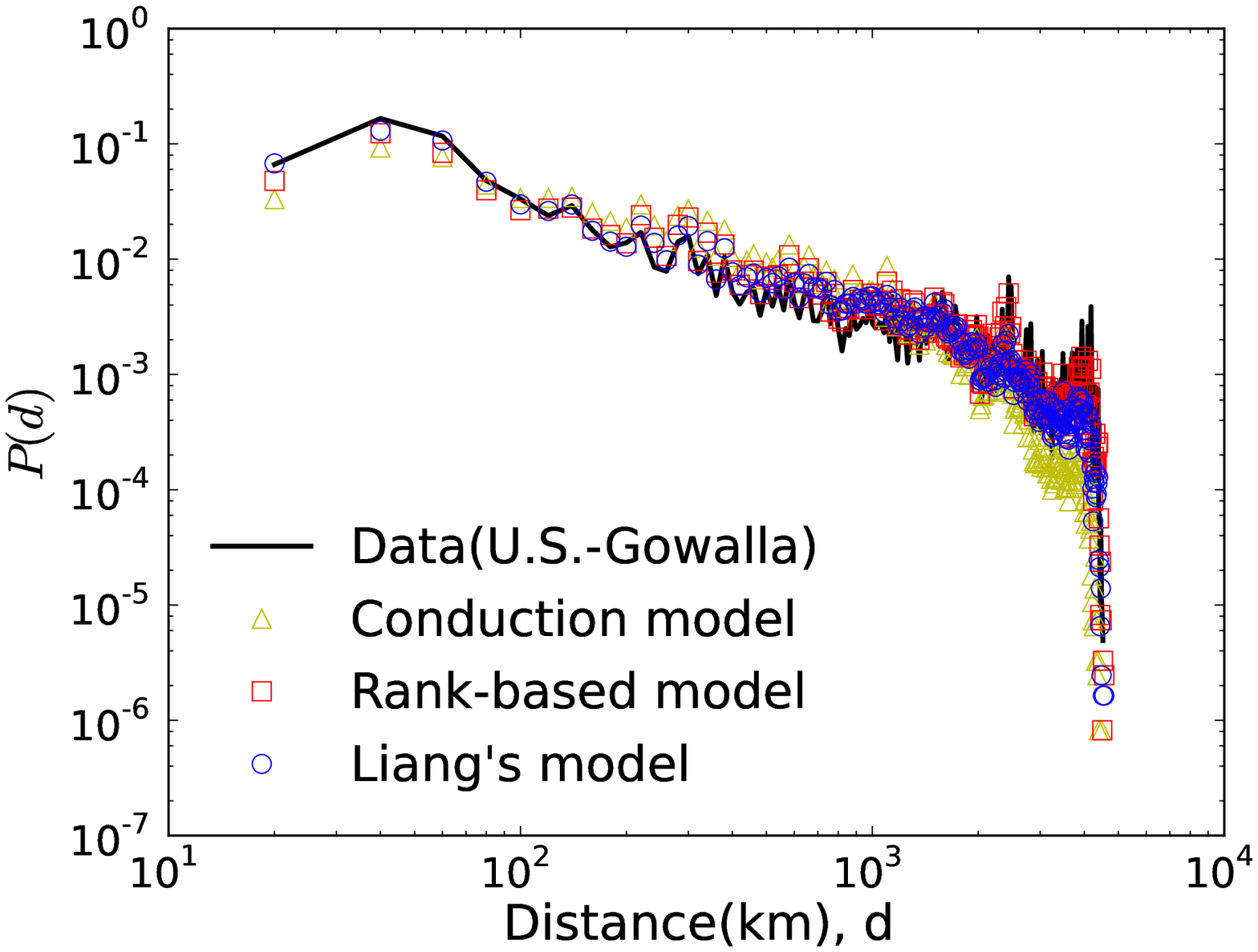}
}
\caption{{\bf The comparison of distance distributions.} (a)Beijing. (b)Chicago. (c)Los Angeles. (d)U.S.(Brightkite). (e)U.S.(Gowalla).}
\label{fig:dist_comp}
\end{figure*}

Here the conduction model, the rank-based model and Liang's model are utilized to simulate human mobility for the datasets, the results of prediction and trip-length distributions are shown in Fig. \ref{fig:flow_comp} and \ref{fig:dist_comp} respectively.

\section*{Social friendship}\label{app:social}
\begin{figure}[htbp]
\centering
\includegraphics[scale=.25]{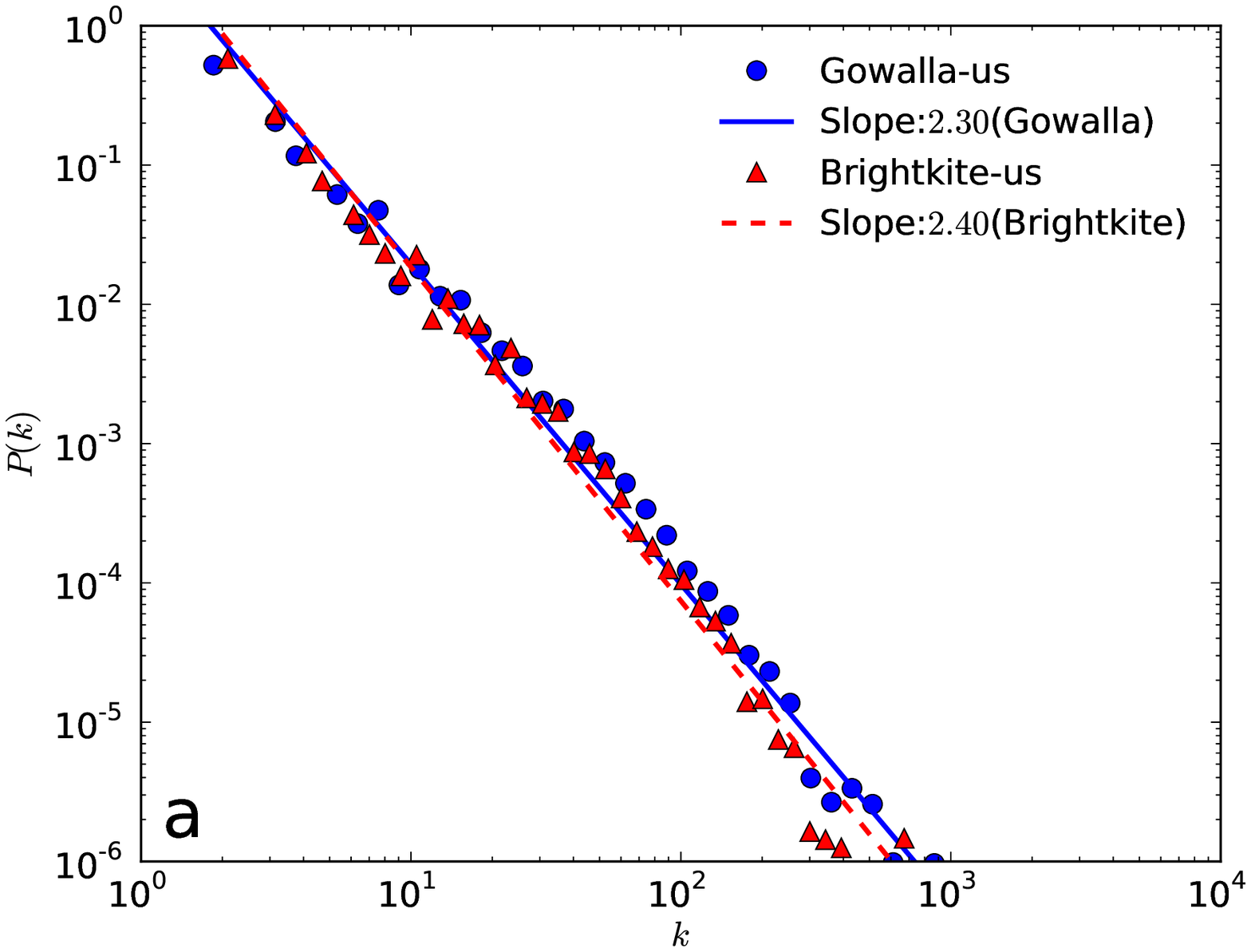}
\includegraphics[scale=.25]{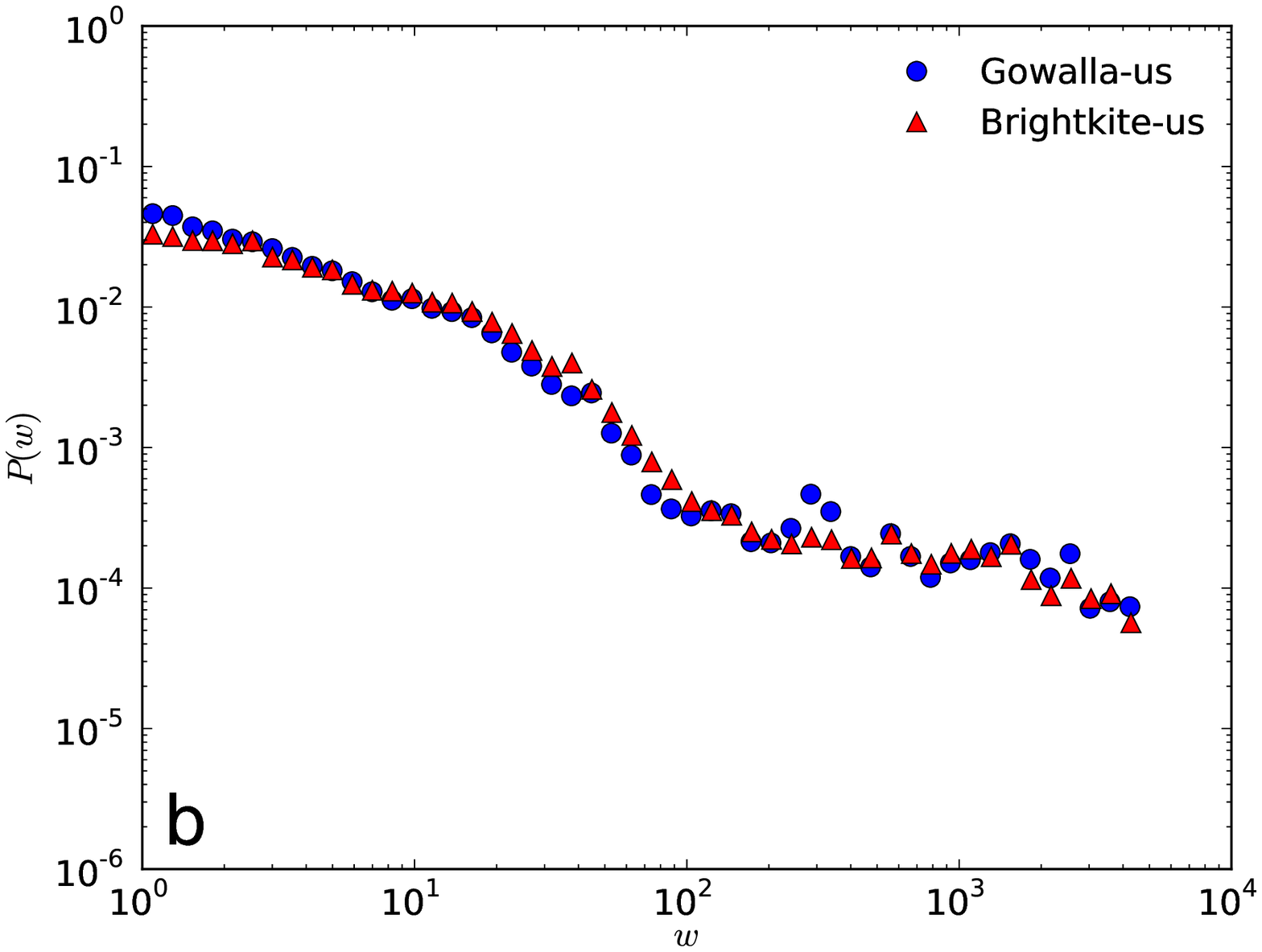}
\caption{{\bf The analysis of the two location-based social networks.} (a)Degree distributions. (b)Weight distributions.}
\label{fig:social_network}
\end{figure}

In the dataset of U.S., the trips were captured by two location-based social services (i.e., Gowalla and Brightkite). During the process of gathering data, besides the location histories of users, the structures of the two social networks were also collected. Intuitively, users often visit more places near their homes resulting in more check-in records. Because of this, the home locations of users can be inferred approximately \cite{Cheng2011}. Then, the weights of edges in both social networks can be defined as the geographic distances between homes of connected users. Consequently, the degree and weight distributions of the two location-based social networks are illustrated in Fig. \ref{fig:social_network}.
From the figure, it can be concluded that both social networks have very similar degree and weight distributions. In Fig. \ref{fig:social_network}(a), both degree distributions like other complex networks are well described by power-law distributions. Moreover, as shown in Fig. \ref{fig:social_network}(b), the probability of being friends drops more and more rapidly with the increase of distance when distance is less than about 100 km, and then decreases slowly. The reason is mainly caused by the heterogeneous geographic distribution of users \cite{Cho2011}. 

\begin{figure}[htbp]
\centering
\includegraphics[scale=.4]{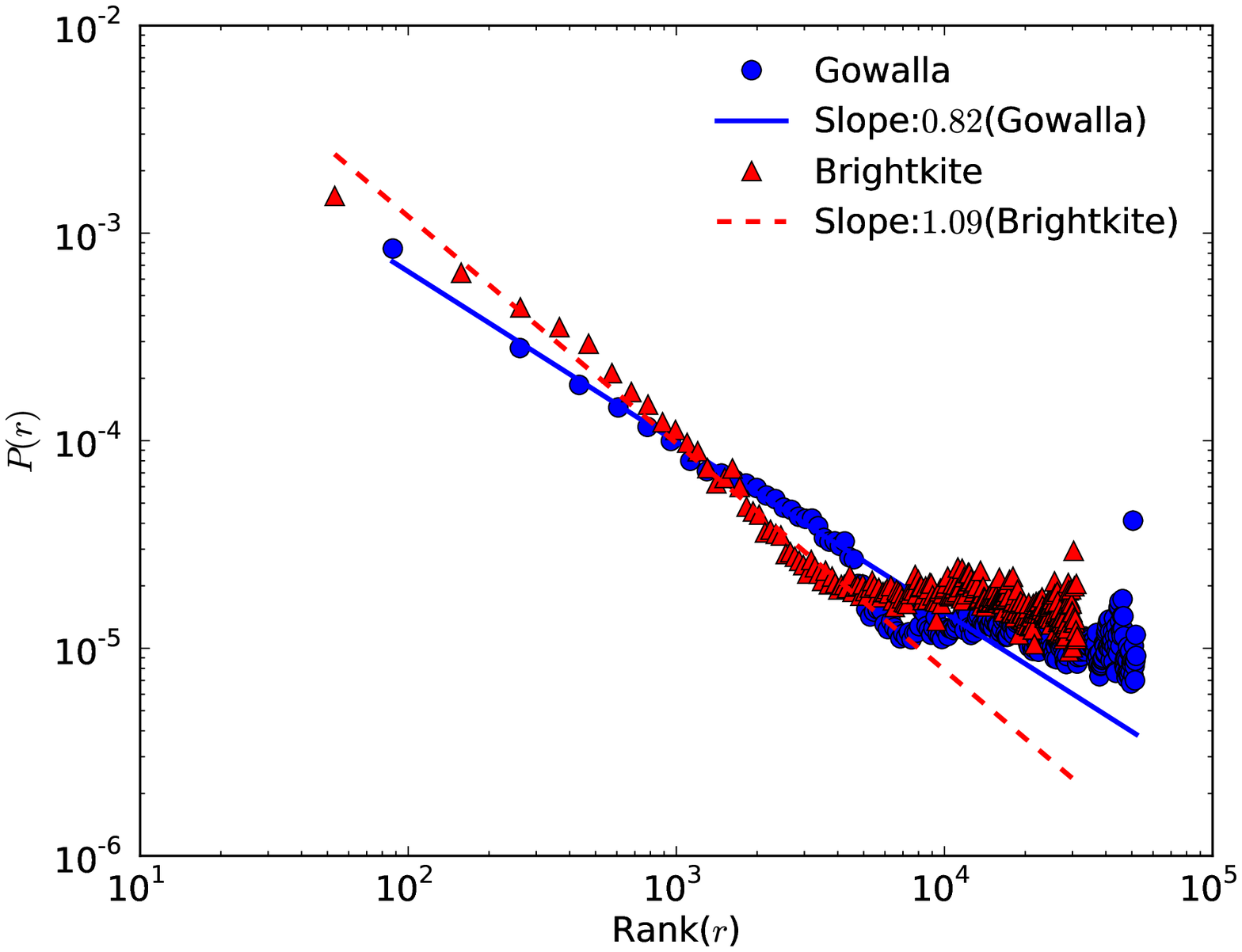}
\caption{{\bf The relationship of probability of making friends versus rank.}}
\label{fig:social_rank}
\end{figure}

Also, we explore the relationship between friendship probability and rank, which is plotted in Fig. \ref{fig:social_rank}. It can be observed that probabilities of making friends in both networks exhibit power-law decays across several orders of magnitude and then become flatten at the tails because of the boundedness of considered geographical environments. The exponents of power law are 0.82 and 1.09 respectively, which are close to the value 1.0 especially for the social network of Brightkite. The property ensures both social networks are navigable effectively without knowing global social structures \cite{Liben-Nowell2005}.







\end{document}